\documentclass[12pt,preprint]{aastex}

\shorttitle{A OBSERVATION TOWARD IRDCs}
\shortauthors{Sakai et al.}

\begin{document}


\title{A MOLECULAR LINE OBSERVATION TOWARD MASSIVE CLUMPS ASSOCIATED WITH INFRARED DARK CLOUDS}


\author{Takeshi Sakai\altaffilmark{1}, Nami Sakai\altaffilmark{2}, Kazuhisa Kamegai\altaffilmark{3}, Tomoya Hirota\altaffilmark{4}, Nobuyuki Yamaguchi\altaffilmark{1}, Shoichi Shiba\altaffilmark{2} and Satoshi Yamamoto\altaffilmark{2}}

\altaffiltext{1}{Nobeyama Radio Observatory, Minamimaki, Minamisaku, Nagano 384-1305, Japan.}
\altaffiltext{2}{Department of Physics, Graduate School of Science, The University of Tokyo, Tokyo 113-0033, Japan.}
\altaffiltext{3}{Institute of Astronomy, The University of Tokyo, Osawa, Mitaka, Tokyo 181-0015, Japan}
\altaffiltext{4}{National Astronomical Observatory of Japan, Osawa, Mitaka, Tokyo 181-8588, Japan.}

\begin{abstract}
We have surveyed the N$_2$H$^+$ $J$=1--0, HC$_3$N $J$=5--4, CCS $J_N$=$4_3$--$3_2$, NH$_3$ ($J$, $K$) = (1, 1), (2, 2), (3, 3), and CH$_3$OH $J$=7--6 lines toward the 55 massive clumps associated with infrared dark clouds by using the Nobeyama Radio Observatory 45 m telescope and the Atacama Submillimeter Telescope Experiment 10 m telescope. The N$_2$H$^+$, HC$_3$N, and NH$_3$ lines are detected toward most of the objects. On the other hand, the CCS emission is detected toward none of the objects. The [CCS]/[N$_2$H$^+$] ratios are found to be mostly lower than unity even in the Spitzer 24 $\mu$m dark objects. This suggests that most of the massive clumps are chemically more evolved than the low-mass starless cores. The CH$_3$OH emission is detected toward 18 out of 55 objects. All the CH$_3$OH-detected objects are associated with the Spitzer 24 $\mu$m sources, suggesting that star formation has already started in all the CH$_3$OH-detected objects.
 The velocity widths of the CH$_3$OH $J_K$=$7_0$--$6_0$ $A^+$ and $7_{-1}$--$6_{-1}$ $E$ lines are broader than those of N$_2$H$^+$ $J$=1--0. 
The CH$_3$OH $J_K$=$7_0$--$6_0$ $A^+$ and $7_{-1}$--$6_{-1}$ $E$ lines tend to have broader linewidth in the MSX dark objects than in the others, the former being younger or less luminous than the latter.
The origin of the broad emission is discussed in terms of the interaction between an outflow and an ambient cloud.
\end{abstract}


\section{Introduction}

Infrared dark clouds (IRDCs) are extinction features against the background mid-IR emission (e.g. Simon et al. 2006), which were recognized by the observations with the mid-infrared satellites such as ISO and MSX (P\'{e}rault et al. 1996; Egan et al. 1998).  
Mapping observations of the mm and sub-mm continuum emissions (Carey et al. 1998; Beuther et al. 2002; Rathborne et al. 2006) revealed that massive clumps are associated with several IRDCs.
Although such massive clumps have a similar mass to the clumps with ongoing high-mass star formation, they are less active.
The gas kinetic temperature of IRDCs is as low as 10--20 K (Carey et al. 1998, Teyssier et al. 2002, Pillai et al. 2006b).
Therefore, massive clumps in IRDCs are thought to be starless or at least in a very early stage of high-mass star formation.

Recently, evidences of high-mass star formation were found in several massive clumps associated with IRDCs. Rathborne et al. (2005) found luminous sources (9,000$\sim$32,000 $L_\odot$) in massive clumps of IRDC G034.43+00.24, which are recognized as high-mass protostars ($\sim$ 10 $M_\odot$). 
Rathborne et al. (2007) also reported that a high-mass protostar ($>$ 8 $M_\odot$) is associated with a massive clump in IRDC G024.33+00.11 on the basis of their high angular resolution mm-wave observation. Beuther et al. (2005b) and Beuther \& Steinacker (2007) found a protostar in IRDC 18223-3 using the Spitzer and the Plateau de Bure data, which would grow up to a high-mass star at the end of its formation process.
In addition to these, indirect evidences of high-mass star formation, such as the CH$_3$OH masers, are found in some IRDCs (Ellingsen 2006; Purcell et al. 2006).

As mentioned above, IRDCs are good targets for studies on high-mass star formation. But little attention has been paid for their chemical compositions, although it would be useful to understand the evolutionary stages of IRDCs.
Since binary chemical reactions, such as neutral-neutral and ion-neutral reactions, are responsible for chemical processes in the deep inside of molecular clumps, the chemical composition changes slowly toward chemical equilibrium with a time scale of $\sim10^6$ yr. This time scale is comparable with the dynamical time scale of molecular cloud clumps. 
Therefore, chemical composition would reflect evolutionary stages of clumps.
For example, carbon-chain molecules, such as CCS and HC$_3$N, are abundant in the early stage of chemical evolution, while N$_2$H$^+$ and NH$_3$ are abundant in the late stage (e.g. Lee et al. 1996).
Furthermore, carbon-chain molecules suffer the great effect from depletion, whereas 
N$_2$H$^+$ and NH$_3$ are less affected by depletion.  When CO, a major destroyer of N$_2$H$^+$, is depleted, N$_2$H$^+$ can be abundant in dense parts. Since NH$_3$ is produced from N$_2$, NH$_3$ can also be abundant (Aikawa et al. 2001, 2005; Bergin \& Tafalla 2007).
Thus, we can investigate the evolutionary stages of clumps from the ratios between carbon-chain molecules and N$_2$H$^+$ or NH$_3$. 

This method has been established in low-mass star forming regions.
Suzuki et al. (1992) and Benson et al. (1998) surveyed the carbon-chain molecule toward
several low-mass cores, and they found that carbon-chain molecules are more abundant in 
starless cores than star-forming cores, whereas NH$_3$ or N$_2$H$^+$ is more abundant in star-forming
cores.  In addition, Hirota et al. (2002, 2004) and Hirota \& Yamamoto (2006) found that 
carbon-chain molecules are very abundant in several low-mass starless cores. 
Recently, Sakai et al. (2007) suggested that this method can also be applied to high-mass star forming regions.  Sakai et al. (2006, 2007) observed the W 3 giant molecular cloud (GMC) with the CCS $J_N$=$4_3$--$3_2$ and N$_2$H$^+$ $J$=1--0 lines, and found that the [CCS]/[N$_2$H$^+$] ratio is higher toward a few massive starless cores in the AFGL 333 cloud than toward the other massive star forming cores in the W 3 GMC.
Sakai et al. (2007) proposed that the massive starless cores in the AFGL 333 cloud are novel candidates which will form high- or intermediate-mass stars, because the CCS emission has hardly been detected in high-mass
star-forming regions (Lai \& Crutcher, 2000).  Note that definitions of "core" and "clump" are described in Sakai et al. (2007).

Chemical composition of clumps would be changed by various activities of newly born stars.  Saturated organic molecules like CH$_3$OH become abundant due to evaporation of grain mantles. The high excitation lines of CH$_3$OH are therefore employed as a tracer of hot cores and shocked regions (e.g. Bachiller \& P\'{e}rez Guti\'{e}rrez 1997).

In this paper, we report a systematic survey of the N$_2$H$^+$ $J$=1--0, CCS $J_N$=$4_3$--$3_2$, HC$_3$N $J$=5--4, CH$_3$OH $J$=7--6 lines, as well as the NH$_3$ ($J$, $K$)=(1, 1), (2, 2) and (3, 3) lines toward massive clumps associated with IRDCs.
Although observations of several molecular lines were carried out toward massive clumps in the IRDCs (Pillai et al. 2006b, 2007; Leurini et al. 2007; Purcell et al. 2006; Ragan et al. 2006; Beuther \& Sridharan 2007), a systematic survey of molecular lines focusing on chemical composition has been limited. In this paper, we present chemical compositions and physical properties of fifty-five nearby massive clumps associated with the IRDCs, and investigate their evolutionary stages.

\section{Observations}

\subsection{The Sample}

Sridharan et al. (2005) identified the candidates of massive starless clumps from the objects found with MAMBO by Beuther et al. (2002). 
Their candidates involve clumps associated with IRDCs.
Rathborne et al. (2006) also observed 38 IRDCs with known kinematic distance by using MAMBO, and listed the positions of clumps.
We prepared a sample of massive clumps associated with IRDCs by use of these two lists on the basis of the following criteria:
\begin{list}{}{}
\item{1.} Distance is less than 4.5 kpc.
\item{2.} Mass is larger than 100 $M_\odot$.
\item{3.} Galactic longitude is less than 34.5 degree.
\end{list}
In addition to these, we also involved the massive clumps located near the clumps listed by Sridharan et al. (2005). Their masses are calculated from the data by Beuther et al. (2002), where the revised equation by Beuther et al. (2005) is employed.
Finally, the sample consists of 55 objects, as listed in Table \ref{tab:t1}.
In the course of this study, 15 objects are eventually found to have the different LSR velocity from those listed by Rathborne et al. (2006).  As the result, such objects do not satisfy the above criteria.

We investigated whether the observed objects are associated with point-like sources or extinction features in the MSX 8 $\mu$m data. Then, 12 objects are found to be associated with the MSX 8 $\mu$m sources, and 26 objects to be associated with extinction features. We call an object with an MSX 8 $\mu$m extinction feature as an MSX dark object.
In the archival Spitzer 24 $\mu$m data, we also inspected whether the observed objects are associated with point-like sources or extinction features of 24 $\mu$m.
As a result, the 31 objects are found to be associated with the point-like 24 $\mu$m sources. 
Note that there are no available Spitzer data for I18151-1208.
There are 11 dark objects in the Spitzer 24 $\mu$m data, which are all MSX 8 $\mu$m dark objects.
These results are summarized in Table \ref{tab:t1}.

\subsection{NRO 45 m observations}

The N$_2$H$^+$ $J$=1--0, CCS $J_N$=$4_3$--$3_2$, HC$_3$N $J$=5--4, NH$_3$ ($J$, $K$) = (1, 1), (2, 2) and (3, 3) lines (Table \ref{tab:t2}) were observed by using the Nobeyama Radio Observatory (NRO) 45 m telescope from January to March in 2007. 
We used the two SIS receivers, S100 for N$_2$H$^+$ and S40 for CCS and HC$_3$N. 
The N$_2$H$^+$, CCS, and HC$_3$N lines were simultaneously observed.
The half-power beam width is about $37^{\prime\prime}$ at 45 GHz and about $18^{\prime\prime}$ at 93 GHz, and the main beam efficiency is 0.75 at 45 GHz and 0.53 at 93 GHz.
The NH$_3$ (1,1), (2,2) and (3, 3) lines were observed with the HEMT receiver. 
The half-power beam width is about $73^{\prime\prime}$ at 22 GHz, and the main beam efficiency is 0.81 at 22 GHz.
For all the observations, acousto-optical radiospectrometers were employed as a backend, each of which has bandwidth and frequency resolution of 40 MHz and 37 kHz, respectively.
The velocity resolutions are 0.50, 0.25, and 0.12 km s$^{-1}$ at 22 GHz, 45 GHz, and 93 GHz, respectively.

The telescope pointing was checked by observing the nearby SiO maser sources every 2 hours, and was maintained to be better than $5^{\prime\prime}$. 
The line intensities were calibrated by the chopper wheel method. 
The system noise temperature was 200--300 K for the N$_2$H$^+$, CCS, and HC$_3$N observations, whereas it was 100--150 K for the NH$_3$ observation.

All the observations were carried out with the position switching mode.
The emission free regions in the Galactic Ring Survey $^{13}$CO $J$=1--0 data (Jackson et al. 2006) were employed as the OFF positions. The data reduction was carried out by use of the AIPS-based software package NewStar developed at NRO.

\subsection{ASTE observations}

The CH$_3$OH $J$=7--6 lines (Table \ref{tab:t2}) were observed with the Atacama Submillimeter Telescope Experiment (ASTE) 10 m telescope (Ezawa et al. 2004; Kohno 2005) from September to October in 2006. We also observed the CH$_3$OH $J_K$=$1_1$--$0_0$ $A^+$ and $J_K$=$4_0$--$3_{-1}$ $E$ lines simultaneously.
The beam size of the ASTE telescope is about 22$^{\prime\prime}$ at 345 GHz.
We used the SIS receiver, SC345, operated in the double-side-band mode.
The backends were autocorrelators, whose band width and resolution each are 512 MHz and 0.5 MHz, respectively.
By using three autocorrelators, we covered the frequency range from 337.9 to 338.8 GHz.
The telescope pointing was calibrated by observing the thermal continuum emission from the Jupiter.
The pointing observation was carried out every observing day, and the pointing accuracy was maintained within 3$^{\prime\prime}$.

All the observations were performed with the position switching mode. 
Antenna temperature was calibrated by the single load chopper-wheel method.
The intensity scale was checked by observing the CH$_3$OH lines toward the NGC 6334 hot core every 0.5--1 hr in order to correct a small variation of the telescope gain due to the temperature and elevation change.
As a result, the relative error in intensity calibration can be maintained to be less than 9 \%.
The main beam efficiency of the ASTE telescope is estimated to be 0.60, although this includes an uncertainty of 20 \%.  The data reduction was carried out by use of the NewStar.

\section{Results}

\subsection{Detection Rates \& Spectra}

Table \ref{tab:t3} shows the 3$\sigma$ detection rates of the observed lines.
The N$_2$H$^+$ $J$=1--0 emission is detected toward 54 out of 55 objects, the HC$_3$N $J$=5--4 emission toward 43 objects, and the NH$_3$ ($J$, $K$) = (1, 1) emission toward all the 55 objects.
On the other hand, the CCS $J_N$=$4_3$--$3_2$ emission is not detected at all.
This survey indicates that the CCS emission is hardly detected in IRDCs, as in the case of high-mass star-forming regions. This is not a distance effect. If the AFGL 333 clump were located on the distance of 4.5 kpc, the peak temperature would be 0.29 K. This value is higher than the upper limits to the CCS peak intensity of all the observed objects, except for I18306-0835 MM1, MM2, and MM3.  Thus, we can detect the CCS line, if a chemically young massive clump like the AFGL 333 clump would exist within the distance of 4.5 kpc. 

The CH$_3$OH $J$=7--6 emissions are detected toward 18 objects.
Although the upper state energy of the CH$_3$OH $J_K$=$7_0$--$6_0$ $A^+$ line is as high as 65 K, the line is detected even in the objects without luminous heating sources seen in the MSX 8 $\mu$m image (MSX dark objects).
In contrast, the Spitzer 24 $\mu$m sources are associated with all the CH$_3$OH-detected objects except for I18151-1208 MM2, for which there are no available Spitzer data.
Thus, it is likely that all the objects with the CH$_3$OH $J$=$7_0$--$6_0$ $A^+$ emission harbor hot regions.

Figure \ref{fig:f1}a shows the spectra of N$_2$H$^+$ $J$=1--0, HC$_3$N $J$=5--4, CH$_3$OH $J_K$=$7_0$--$6_0$ $A^+$, and CCS $J_N$=$4_3$--$3_2$ for the selected objects, whereas Figure \ref{fig:f2}a shows those of NH$_3$ ($J$, $K$) = (1, 1), (2, 2) and (3, 3). The spectra in Figures \ref{fig:f1}a and \ref{fig:f2}a are "typical ones" taken arbitrarily from the spectra of our sample.
The observed spectra toward all the objects are shown in Figures \ref{fig:f1}b--e and \ref{fig:f2}b--e, which are available in the electric edition of the Journal.
Note that the observed lines toward G024.08+00.04 MM2 have two velocity components (52 km s$^{-1}$ and 114 km s$^{-1}$).
The N$_2$H$^+$ $J$=1--0 line shows complicated hyperfine splittings due to the nuclear spin ($I$=1) of the nitrogen nuclei. 
The observed line shows triplet structure, where the center and higher velocity components consist of three hyperfine lines each unresolved due to the velocity width.
The velocity width, optical depth, and excitation temperature of N$_2$H$^+$ are derived by fitting the observed spectrum pattern to a multi Gaussian function (see Appendix B in Sakai et al. 2006).
The results are given in Table \ref{tab:t4}.
We fit the spectra of the HC$_3$N and CH$_3$OH lines to a single Gaussian function, and these results are summarized in Tables \ref{tab:t5} and \ref{tab:t6}, respectively.
We represent the upper limits to peak temperature and integrated intensity for the non-detected lines in the tables.
The upper limits to peak temperature for the non-detected lines are a 3$\sigma$ value at the velocity resolution of one channel width. 
The upper limits to integrated intensity for the non-detected lines are derived by using the following equation;
$3 T_{\rm rms} \sqrt{10 ({\rm km s^{-1}}) \Delta V_{\rm res}}$, where 
$\Delta V_{\rm res}$ is a channel width in km s$^{-1}$, and $T_{\rm rms}$ is a rms noise level at the velocity resolution of $\Delta V_{\rm res}$.

As for the NH$_3$ (1, 1) line, five hyperfine components are fitted to the single Gaussian each, and the parameters of the main hyperfine components are listed in Table \ref{tab:t7}.
On the other hand, the NH$_3$ (2, 2) and (3, 3) lines appear as a single line without hyperfine structure, and hence, they are fitted to a single Gaussian function each (Tables \ref{tab:t8} and \ref{tab:t9}).
Note that the NH$_3$ (1, 1) and (2, 2) lines show a double-peak feature in G024.33+00.11 MM3 and MM9. 
Since the double peak structure is seen even in the weak hyperfine components, it seems to be a velocity structure.
Then, we fit the spectra to a double Gaussian function.
The parameters for each peak are listed in Tables \ref{tab:t7} and \ref{tab:t8}.

As mentioned earlier, we found that the LSR velocities of 15 objects including G024.33+00.11 are much different from the values in the lists of Rathborne et al. (2006). 
This is because they estimated the systemic velocity of the objects from the $^{13}$CO $J$=1--0 line.
The critical density of N$_2$H$^+$ $J$=1--0 is much higher than that of $^{13}$CO $J$=1--0, and hence, the N$_2$H$^+$ $J$=1--0 emission well traces the dense clumps. 
We re-evaluate the kinematic distance of the dense clumps from the LSR velocity of the N$_2$H$^+$ or NH$_3$ lines for all the observed objects by using the rotation curve obtained by Clemens (1985) with ($\Theta_0$, $R_0$) = (220 km s$^{-1}$, 8.5 kpc).  The derived distances are listed in Table \ref{tab:t4}.

\begin{figure}
\figurenum{3}
 \epsscale{1.1}
 \plotone{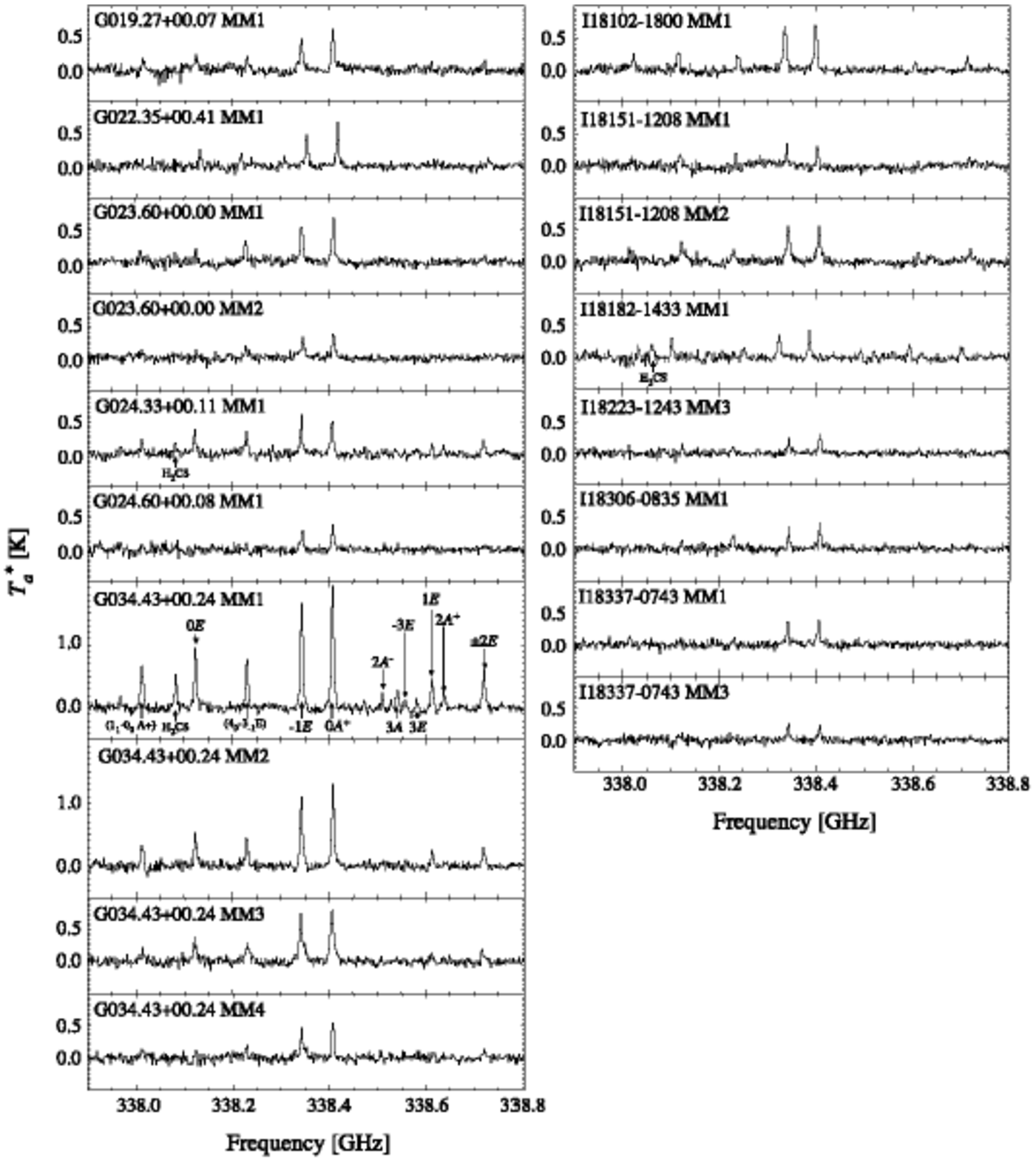}
	 \caption{Spectra of the CH$_3$OH $J$=7--6 lines.}
	\label{fig:f3}
\end{figure}

Figure \ref{fig:f3} shows the spectra covering the CH$_3$OH $J_K$=$7_K$--$6_K$ transitions toward the CH$_3$OH-detected objects. 
Two strong lines are seen around 338.4 GHz; these are the $J_K$=$7_{-1}$--$6_{-1}$ $E$ and $J_K$=$7_0$--$6_0$ $A^+$ lines.
In addition, higher excitation lines ($7_0$--$6_0$ $E$ and $7_{\pm 2}$--$6_{\pm 2}$ $E$) are seen toward most of the sources.
The CH$_3$OH $J_K$=$1_1$--$0_0$ $A^+$ and $J_K$=$4_0$--$3_{-1}$ $E$ lines from the other side band are also observed.
Although these two lines are lower-$J$ lines, the intensities of these lines are 
generally lower than those of the $J_K$=$7_0$--$6_0$ line because of their lower line strength.
Therefore, the two lower-$J$ CH$_3$OH lines are not detected toward the objects without 
CH$_3$OH $J_K$=$7_0$--$6_0$ $A^+$ emission.  The strongest CH$_3$OH source is G034.43+00.24 MM1, where even the $J_K$=$7_3$--$6_3$ lines ($E_u$ = 112--127 K) are detected. 
The H$_2$CS 10$_{1,10}$--9$_{1,9}$ line ($E_u$ = 102 K) is detected toward three objects, G024.33+00.11 MM1, G034.43+00.24 MM1, and I18182-1433 MM1, which seem to contain hot core activities.
In fact, Rathborne et al. (2007) reported the detection of many molecular lines, such as HCOOCH$_3$ and CH$_3$CH$_2$CN, toward G024.33+00.11 MM1.
Since large organic molecules are thought to be formed from molecules evaporated from dust grains, such as C$_2$H$_2$, HCN, and CH$_3$OH, in the hot gas phase (e.g. Nomura \& Millar, 2004), the detections of HCOOCH$_3$ and CH$_3$CH$_2$CN definitively indicate the hot core activity.

\subsection{Integrated Intensities}

\begin{figure}
\figurenum{4}
 \epsscale{1.0}
 \plotone{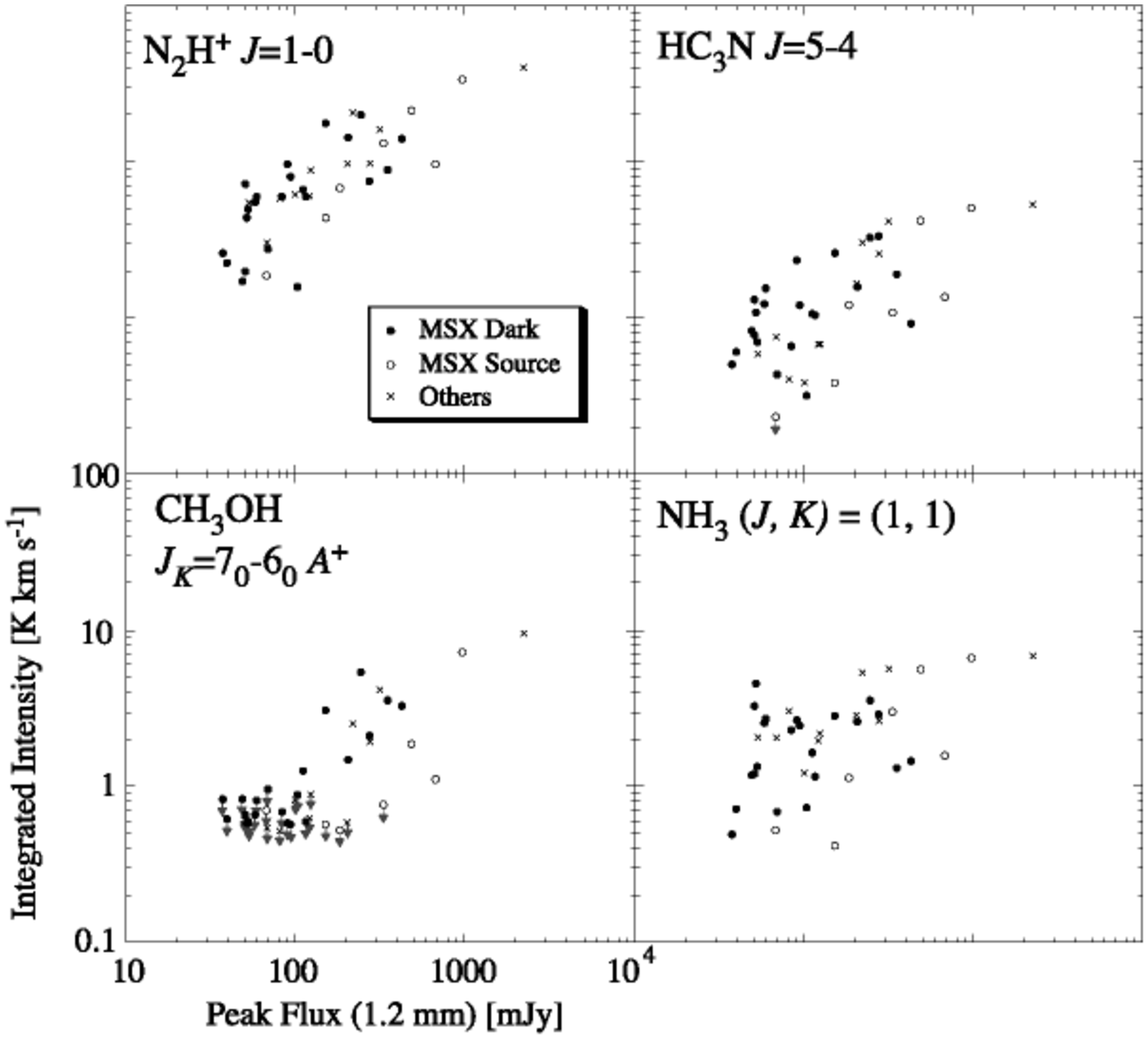}
	 \caption{Correlation plots of the integrated intensities against the 1.2 mm continuum peak flux. The 1.2 mm continuum data are observed by Beuther et al. (2002) and Rathborne et al. (2006). The values of the objects with $D$ $\leq$ 4.5 kpc are plotted.}
	\label{fig:f4}
\end{figure}

In Figure \ref{fig:f4}, we plot the integrated intensities of a few observed lines against the peak flux of the 1.2 mm continuum, $F_{1.2}$, reported by Beuther et al. (2002) and Rathborne et al. (2006).
We plot only the values of the objects with $D$ $\leq$ 4.5 kpc.
The mm-wave dust continuum emission is optically thin in general, and it is a good tracer of column density of a dense region. 
A correlation between the N$_2$H$^+$ $J$=1--0 emission and the dust continuum emission was suggested in several regions (Caselli et al. 1999, 2002a; Bergin et al. 2001; Sakai et al. 2007), although it was also reported that the N$_2$H$^+$ emission does not always trace the structures by dust continuum emission in very active star forming regions, such as Orion KL (Ungerechts et al. 1998).
In IRDCs, the N$_2$H$^+$ emission is correlated with the 1.2 mm peak flux, as shown in Figure \ref{fig:f4}, confirming that the N$_2$H$^+$ emission well traces the column density of a dense regions. 
The HC$_3$N integrated intensity is also roughly correlated with the 1.2 mm peak flux, suggesting that the HC$_3$N emission also comes from a dense region, as well as the N$_2$H$^+$ emission.
The CH$_3$OH $J_K$=$7_0$--$6_0$ $A^+$ line is only detected toward the objects where $F_{1.2}$ is higher than 100 mJy.
By using the relation reported by Beuther et al. (2005a), $F_{1.2}$ of 100 mJy corresponds to $A_V$ of 49, 68, and 114 at $T_k$ of 40, 30, and 20 K, respectively.
Furthermore, all the objects with $F_{1.2}$ $>$ 130 mJy are the Spitzer sources.
Hence, the appearance of the CH$_3$OH $J$=7--6 emission is related to the star formation activity.

A marginal correlation can be seen between the NH$_3$ integrated intensities and the 1.2 mm peak flux. The NH$_3$ integrated intensities are scattered even at the low $F_{1.2}$ values, probably because the NH$_3$ emission traces less dense gas due to its lower critical density ($\sim$ 10$^3$ cm$^{-3}$), and also because the beam size of the NH$_3$ observation (73$^{\prime\prime}$) is larger than those of the other observations (18$^{\prime\prime}$--37$^{\prime\prime}$).

\section{Discussions}

\subsection{Temperatures}\label{sec:temp}

We derive the NH$_3$ rotation temperature, $T_{\rm rot}$(NH$_3$), from the NH$_3$ (1, 1) and (2, 2) lines by using the method presented by Ho \& Townes (1983) and Li et al. (2003).
We first evaluate the optical depth of the NH$_3$ (1, 1) line from the peak intensity ratio of the main and satellite hyperfine components. Then we calculate the rotation temperature from the peak intensity ratio of the main hyperfine components of the (1, 1) and (2, 2) lines considering the optical depth of the (1, 1) line obtained above. 
The derived $T_{\rm rot}$(NH$_3$) values are listed in Table \ref{tab:t10}. 
A range of $T_{\rm rot}$(NH$_3$) is from 10.3 to 20.8 K with the median of 14--16 K. This is higher than the corresponding values of $T_{\rm rot}$(NH$_3$) reported for dark clouds  (10--15 K; Benson \& Myers, 1989).

\begin{figure}
\figurenum{5}
 \epsscale{1.0}
 \plotone{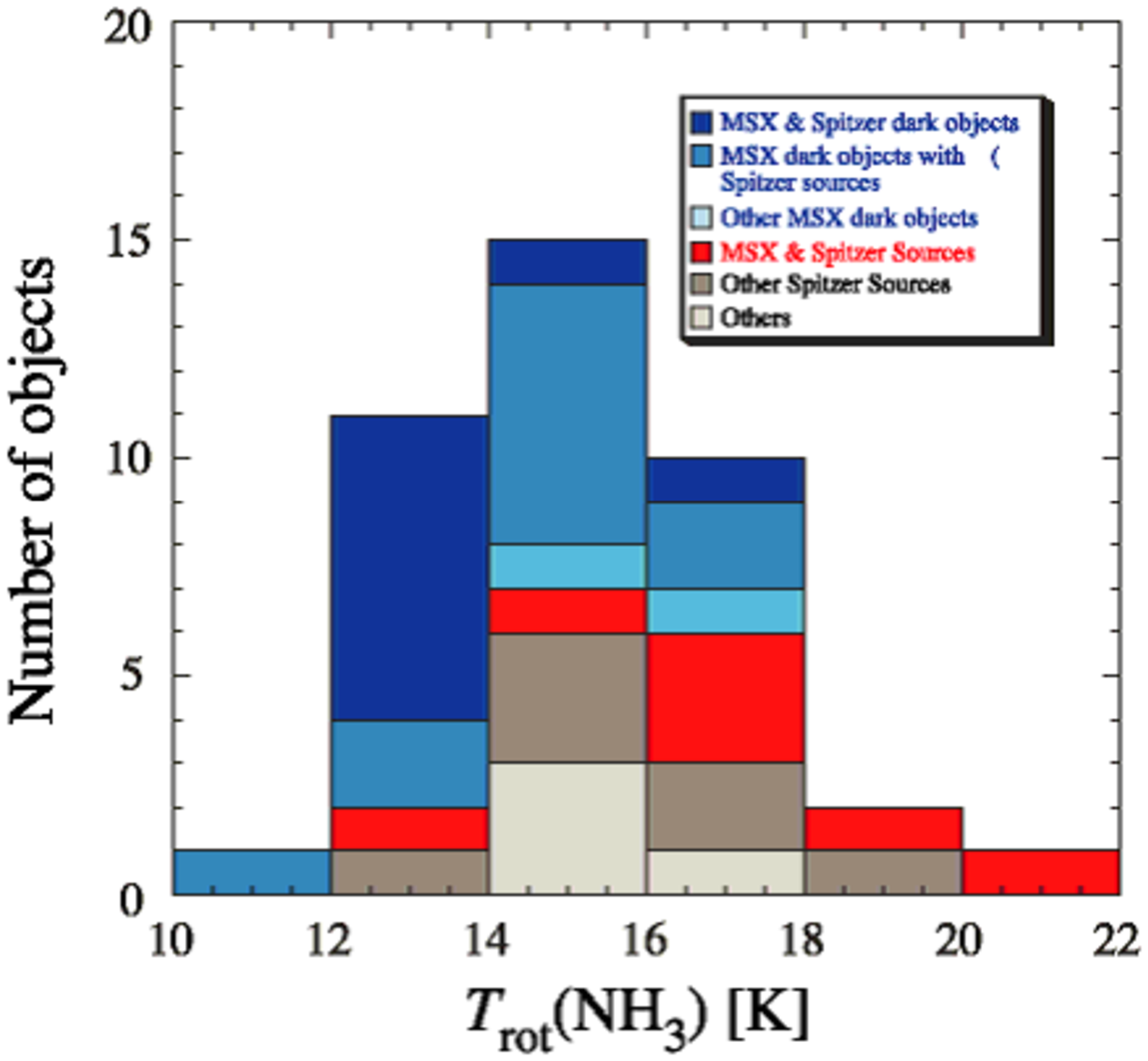}
	 \caption{Histogram of the NH$_3$ rotation temperature. The objects with $D$ $\leq$ 4.5 kpc are used for the histogram.}
	\label{fig:f5}
\end{figure}

Figure \ref{fig:f5} represents a histogram of $T_{\rm rot}$(NH$_3$) for the objects with $D$ $\leq$ 4.5 kpc.
Figure \ref{fig:f5} shows that $T_{\rm rot}$(NH$_3$) for the MSX sources tends to be higher than that for the MSX dark objects.
This confirms that the emitting regions are associated with the MSX sources. 
The $T_{\rm rot}$(NH$_3$) values of the MSX and Spitzer dark objects tend to be lower than those of the other objects.
The average $T_{\rm rot}$(NH$_3$) value of the MSX and Spitzer dark objects is 13.9$\pm$1.5 K, which is comparable to that for the other samples of the MSX-dark IRDCs (13.9 K: Pillai et al. 2006b; 15.3 K:  Sridharan et al. 2005), but is lower than that for the high-mass protostellar objects (22.5 K: Sridharan et al. 2002). 

The rotation temperatures of CH$_3$OH, $T_{\rm rot}$(CH$_3$OH), are derived by using the rotation diagram method (e.g. Turner 1991) under the assumption of local thermodynamic equilibrium (LTE). Since several E type lines are detected in many objects (Figure \ref{fig:f3}), we derive the rotation temperature from them.
The results are listed in Table \ref{tab:t11}.
The derived $T_{\rm rot}$(CH$_3$OH) values range from 7 to 12 K.
The range is lower than that for the high-mass star forming cores (24--203 K reported by van der Tak et al. (2000b), if we exclude one exception of GL 7009S (8 K)).
The average value of $T_{\rm rot}$(CH$_3$OH) in the MSX dark objects is 8.0$\pm$0.9 K, which seems to be slightly lower than that of the MSX sources (10$\pm$2 K).
Although the upper state energy of the CH$_3$OH $J$=$7_{-1}$--$6_{-1}$ $E$ line is 70 K,
the derived temperatures are less than 12 K, which are even lower than $T_{\rm rot}$(NH$_3$). This may suggest that the rotational levels of CH$_3$OH are not thermalized to the gas kinetic temperature.  Leurini et al. (2007) modeled the CH$_3$OH emission toward several IRDCs, and they derived the gas kinetic temperatures for an extended component, an outflow, and a core to be 24$\pm$7, 15$\pm$5, and 47$\pm$9 K, respectively.
The $T_{\rm rot}$(CH$_3$OH) values derived in the present study are lower than their values, implying that the CH$_3$OH lines are subthermally excited.

\subsection{Column Densities}

We derive the column densities of the observed molecules. For simplicity, we assume the LTE condition for all the molecules.
As for N$_2$H$^+$, we derive the column density by assuming that the excitation temperature is equal to the rotation temperature of NH$_3$, and the lines are optically thin. 
This can be justified, because $\Delta V$(N$_2$H$^+$) is similar to $\Delta V$(NH$_3$), as mentioned in the next section. 
The derived column densities are listed in Table \ref{tab:t10}.
In the case of G034.43+00.24 MM1, a change in $T_{\rm ex}$ by $\pm$1 K results in a change in the derived column density by $\pm$4 \%.
Most of the N$_2$H$^+$ column densities are found to be larger than 1$\times$10$^{13}$ cm$^{-2}$.
This is higher than a typical value found in low mass cores, which is several times 10$^{12}$ cm$^{-2}$ (Benson et al. 1998; Caselli et al. 2002b). 

The upper limit to the CCS column density is derived by assuming the optically thin condition.
The rotation temperature of NH$_3$ is employed as the excitation temperature.
The derived upper limits are listed in Table \ref{tab:t10}.
They are found to be lower than 7$\times$10$^{12}$ cm$^{-2}$ for all the objects.
These are lower than the CCS column densities of some low-mass starless cores in the dark clouds, typically a few times 10$^{13}$ cm$^{-2}$ (Suzuki et al. 1992; Benson et al. 1998), and are also lower than the CCS column density of the massive clump in the AFGL 333 cloud, which is (1.2--3.0)$\times$10$^{13}$ cm$^{-2}$ (Sakai et al. 2007).
Table \ref{tab:t10} lists the upper limit to the [CCS]/[N$_2$H$^+$] ratio.
Most of the ratios are lower than 1 even in the Spitzer dark objects,
whereas the corresponding ratio is as high as 2.6--3.2 for some low-mass starless cores (Benson et al. 1998).
Since we use the [CCS]/[N$_2$H$^+$] ratio to discuss the chemical evolution, the distance effect would be cancelled.
It is therefore suggested that the massive clumps associated with the IRDCs are chemically evolved than low mass starless cores.
This survey demonstrates that the AFGL 333 clump could be a rare case of a chemically young massive clump. 

The HC$_3$N column density is evaluated by assuming the optically thin condition, where the excitation temperature is assumed to be equal to $T_{\rm rot}$(NH$_3$). 
In the case of G034.43+00.24 MM1, a change in $T_{\rm ex}$ by $\pm$1 K results in a change in the derived column density by $\pm$3 \%.
In contrast to CCS, the derived HC$_3$N column densities (Table \ref{tab:t10}) are comparable to those for the low-mass starless cores, $\sim$10$^{13-14}$ cm$^{-2}$(Suzuki et al. 1992), even in the objects with MSX sources, probably because HC$_3$N is less sensitive to chemical evolution than CCS.
In addition, HC$_3$N could also be formed in a hot gas. In fact, strong HC$_3$N emission is detected toward hot core sources like Orion KL (Ungerechts et al. 1998). 

The NH$_3$ column density is derived from the optical depth and rotation temperature (Section \ref{sec:temp}).
The derived column densities are listed in Table \ref{tab:t10}.
The NH$_3$ column densities (a few times 10$^{15}$ cm$^{-2}$) are higher than those of the low mass cores, $\sim$10$^{14-15}$ cm$^{-2}$ (Suzuki et al. 1992), and are comparable to those of the IRDCs observed by Pillai et al. (2006b).
The $N$(CCS)/$N$(NH$_3$) ratios are found to be lower than 0.007 for most cases, which are lower than those in some low-mass dark cloud cores ($>$0.01) (Suzuki et al. 1992).
Along with the results of the $N$(CCS)/$N$(N$_2$H$^+$) ratio, this further supports that the observed IRDCs are chemically evolved.

We derive the CH$_3$OH column density from the integrated intensity of the CH$_3$OH $J_K$=$7_0$--$6_0$ $A^+$ line. In this calculation, $T_{\rm rot}$(CH$_3$OH) is not used as the excitation temperature, because the rotational levels of CH$_3$OH are not thermalized as mentioned before.
$T_{\rm rot}$(CH$_3$OH) is an effective temperature determined by the population difference between the different $K$ ladders, and does not always represent the excitation temperature of the $7_0$--$6_0$ $A^+$ line. Therefore, we assume the excitation temperature ranging from 20 K to 50 K by referring the result by Leurini et al. (2007).
We also estimate the upper limit to the CH$_3$OH column density for the non-detected objects with adopting the rotation temperature of 20 K.
The derived CH$_3$OH column densities ranges from 2.9$\times$10$^{14}$ to 4.9$\times$10$^{15}$ cm$^{-2}$ toward the CH$_3$OH-detected objects, as shown in Table \ref{tab:t10}. 
They are comparable to those in the IRDCs obtained by Leurini et al. (2007), although they tend to be lower than those in several IRDCs obtained by Beuther \& Sridharan (2007) (1.7$\times$10$^{13}$--3.7$\times$10$^{14}$ cm$^{-2}$).
In addition, they are also comparable to those in the massive star forming regions obtained by van der Tak et al. (2000b). 
On the other hand, Maret et al. (2005) reported that the CH$_3$OH column densities ranges from 5.3$\times$10$^{13}$ to 8.1$\times$10$^{14}$ cm$^{-2}$ in the low-mass star forming regions, NGC 1333, L1448, L1157, and IRAS16293-2422.
Thus, the CH$_3$OH column densities tend to be higher than those in the low-mass star forming regions.

\subsection{Velocity Widths \& Star Formation}

\begin{figure}
\figurenum{6}
 \epsscale{1.0}
 \plotone{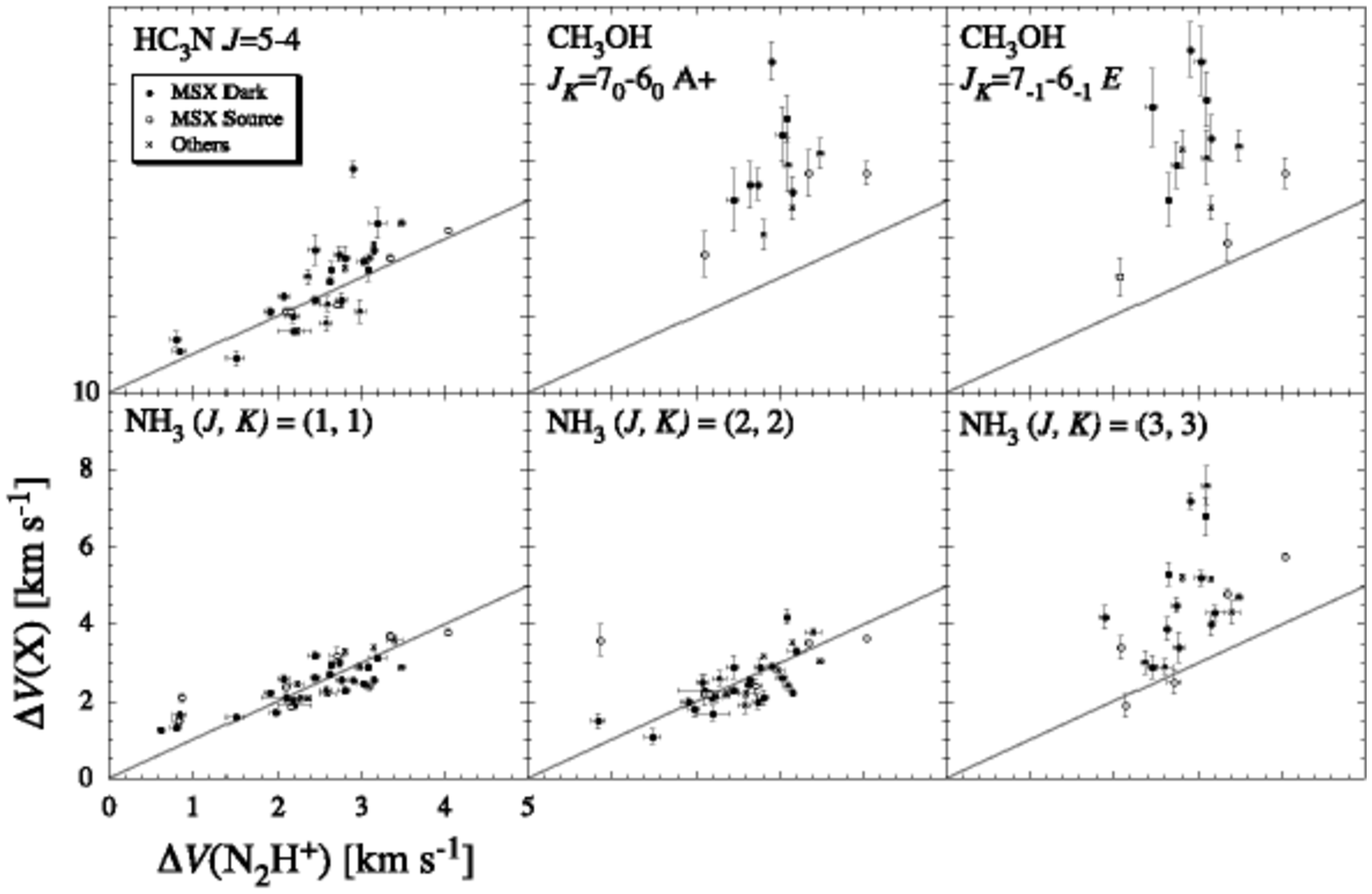}
	 \caption{Correlation plots of the N$_2$H$^+$ $J$=1--0 velocity width against the velocity widths of HC$_3$N $J$=5--4, CH$_3$OH $J_K$=$7_0$--$6_0$ $A^+$, CH$_3$OH $J_K$=$7_{-1}$--$6_{-1}$ $E$, NH$_3$ ($J$, $K$) = (1, 1), (2, 2) and (3, 3). The objects with $D$ $\leq$ 4.5 kpc are plotted.}
	\label{fig:f6}
\end{figure}

\begin{figure}
\figurenum{7}
 \epsscale{1.0}
 \plotone{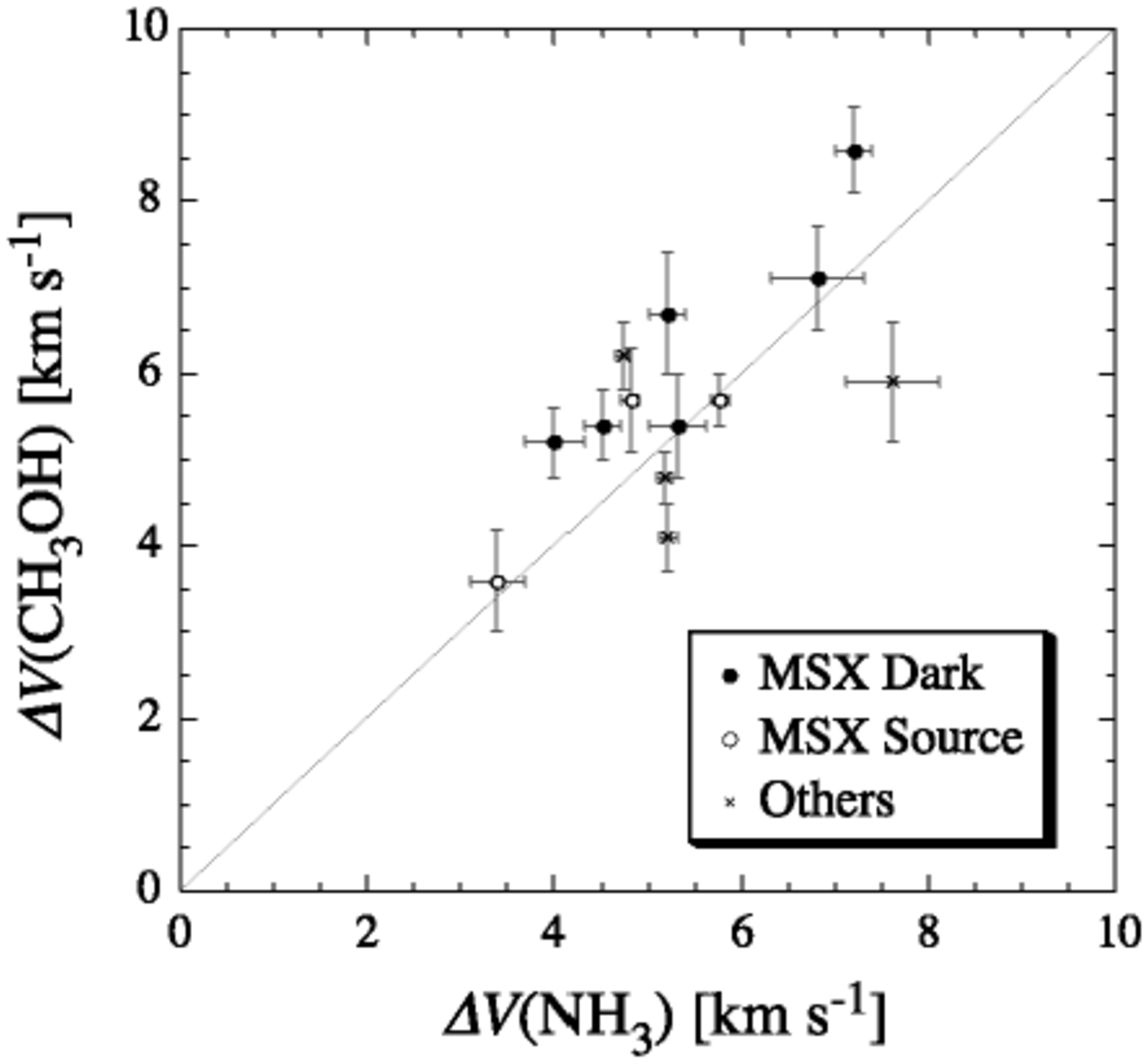}
	 \caption{Correlation plots of the velocity width of CH$_3$OH $J_K$=$7_0$--$6_0$ $A^+$ against the velocity width of NH$_3$ ($J$, $K$) = (3, 3). The objects with $D$ $\leq$ 4.5 kpc are plotted.}
	\label{fig:f7}
\end{figure}

Velocity widths give information about turbulence of clumps including effects of star formation activity.
In Figure \ref{fig:f6}, we plot the velocity widths of the HC$_3$N $J$=5--4, CH$_3$OH $J_K$=$7_0$--$6_0$ $A^+$, NH$_3$ ($J$, $K$) = (1, 1), (2, 2) and (3, 3) lines against that of the N$_2$H$^+$ $J$=1--0 line for the objects with $D$ $\leq$ 4.5 kpc.
The velocity widths of the HC$_3$N, NH$_3$ (1, 1) and (2, 2) lines are correlated with that of N$_2$H$^+$, indicating that they come from almost the same region. On the other hand, the correlation is poor for high excitation lines such as CH$_3$OH ($J_K$=$7_0$--$6_0$ $A^+$ and $7_{-1}$--$6_{-1}$ $E$) and NH$_3$ (3, 3).
Since the upper state energies of CH$_3$OH $J$=$7_K$--$6_K$ (65 K) is higher than that of N$_2$H$^+$ $J$=1--0 (4 K) and that of NH$_3$ (3, 3) is as high as 124.5 K, the CH$_3$OH $J$=7--6 and NH$_3$ (3, 3) lines would trace the denser and hotter region near the protostar.
In fact, the velocity widths of the CH$_3$OH $J$=7--6 and NH$_3$ (3, 3) lines are correlated with each other (Figure \ref{fig:f7}).
Pillai et al. (2006a) also found that the NH$_3$ (3, 3) velocity width is larger than the NH$_3$ (1, 1) and (2, 2) velocity widths toward the IRDC, G11.11-0.12.
Beuther et al. (2002) reported a similar trend in the C$^{34}$S and NH$_3$ velocity widths, where the C$^{34}$S $J$=2--1 velocity widths are about 1.5 times larger than the NH$_3$ (1, 1) velocity widths. 
Since the C$^{34}$S emission is optically thin and the critical density of C$^{34}$S is as high as $3\times10^5$ cm$^{-3}$, it also traces dense regions. Furthermore, CS could be abundant in the 
hot gas phase in contrast to N$_2$H$^+$ (e.g. Nomura \& Millar 2004). Thus it is likely that the 
C$^{34}$S emission comes from regions close to the newly born stars, as well as the CH$_3$OH and NH$_3$ (3, 3) emissions.

\begin{figure}
\figurenum{8}
 \epsscale{0.5}
 \plotone{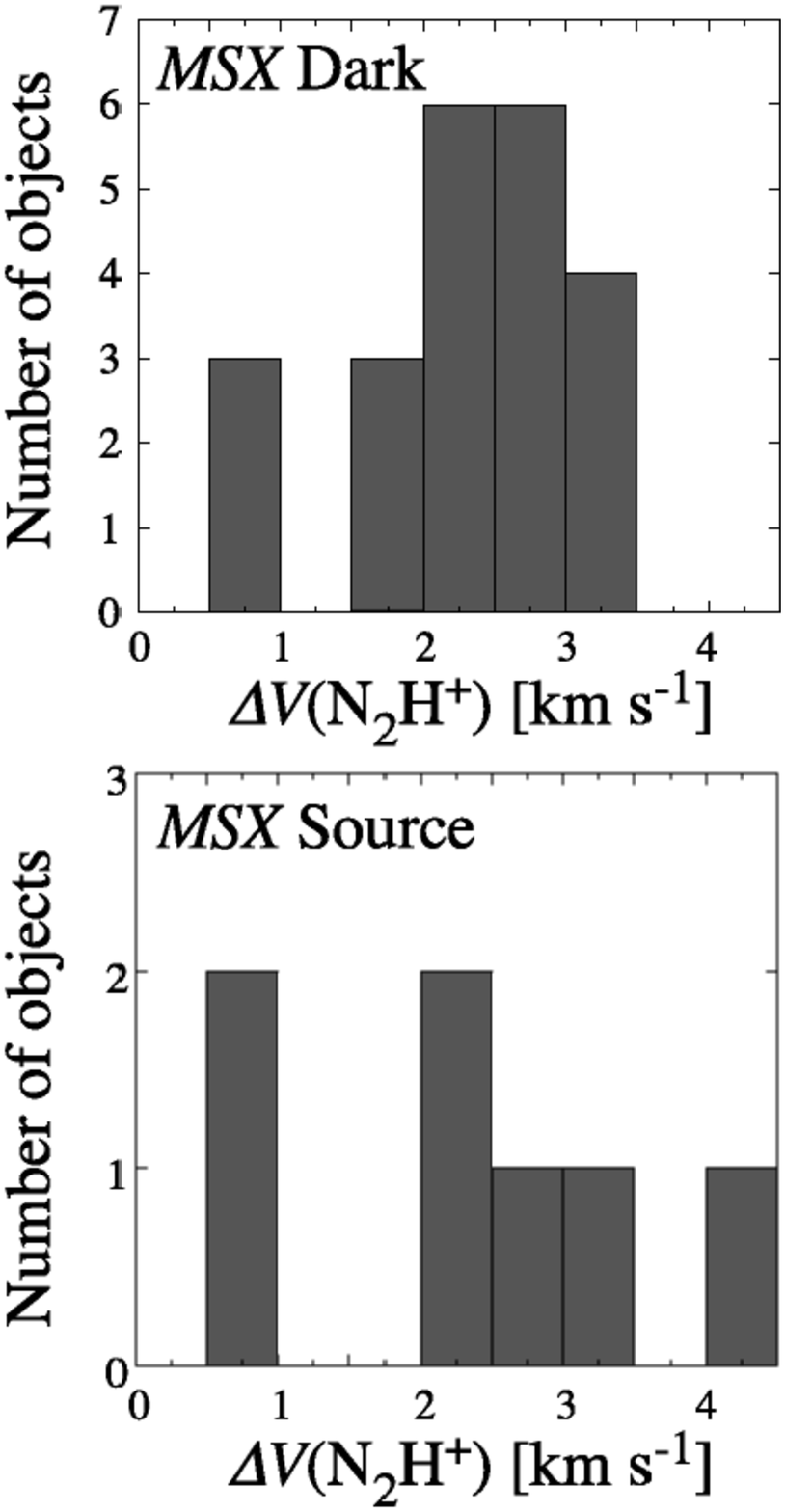}
	 \caption{Histograms of the N$_2$H$^+$ velocity width of the MSX dark objects (top) and MSX sources (bottom). The objects with $D$ $\leq$ 4.5 kpc are plotted.}
	\label{fig:f8}
\end{figure}

Figure \ref{fig:f8} shows histograms of the velocity width of N$_2$H$^+$ for the MSX dark objects and the objects with MSX sources. 
The mean N$_2$H$^+$ velocity width of the MSX dark objects (2.3 km s$^{-1}$) is almost comparable to that of the objects with MSX sources (2.5 km s$^{-1}$).
These values are also comparable to that of the high-mass star forming regions observed by Pirogov et al. (2003; 2.4 km s$^{-1}$).
The histogram shapes are also similar to each other. 
Since the N$_2$H$^+$ emission arises from the whole clump, it is not sensitive to the star formation activities. 
The mean N$_2$H$^+$ velocity widths of the MSX and Spitzer dark objects (2.1 km s$^{-1}$) are significantly larger than those of the low-mass starless cores ($\sim$0.5 km s$^{-1}$) (Benson et al. 1998), and are also larger than those of the cores in a giant molecular cloud, Orion A (Tatematsu et al. 1993). 
The massive clumps associated with the IRDCs seem to be initially turbulent.
This gives an important constraint when one would construct a model of star-formation processes in IRDCs.
For instance, McKee \& Tan (2003) argued that massive stars can form in turbulence-supported 
cores with a high accretion rate.  Thus, the large velocity widths observed in IRDCs are consistent with the theoretical prediction, if the IRDCs are birthplaces of high-mass stars.

\begin{figure}
\figurenum{9}
 \epsscale{0.8}
 \plotone{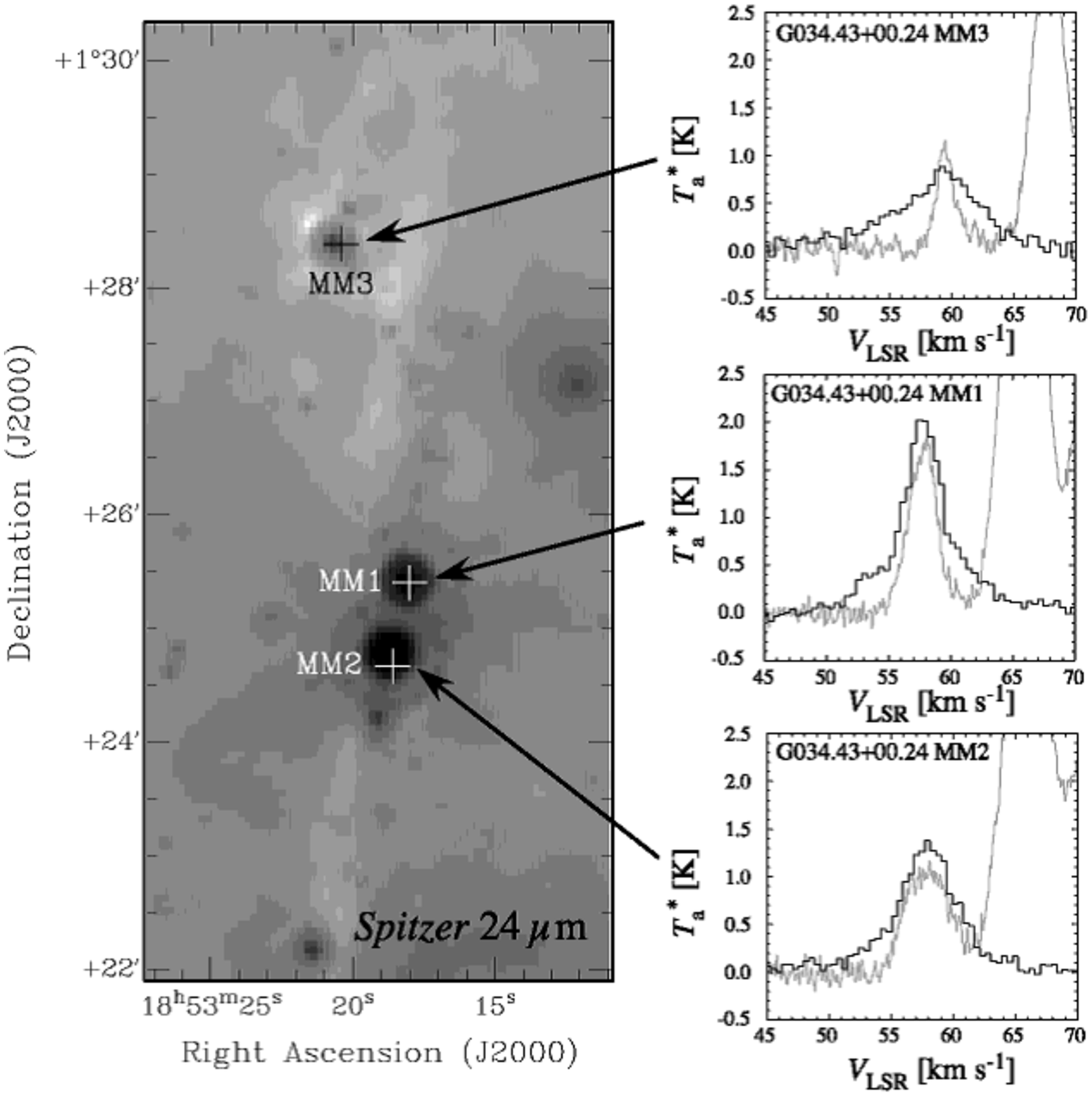}
	 \caption{Spizter 24 $\mu$m image toward G034.43+00.24 (left), and spectra of the CH$_3$OH $J_K$=$7_0$--$6_0$ $A^+$ (black) and N$_2$H$^+$ $J$=1--0 $F_1$=0--1 $F$=1--2 lines (gray) toward G034.43+00.24 MM1, MM2 and MM3 (right).}
	\label{fig:f9}
\end{figure}

As shown in Figure \ref{fig:f6}, the CH$_3$OH $J$=7--6 velocity widths for the MSX dark objects tend to be broader than those for the objects with MSX sources at a given N$_2$H$^+$ velocity width.
This trend is clear in the case of G034.43+00.24. 
Figure \ref{fig:f9} shows the Spitzer 24 $\mu$m image toward the G034.43+00.24 region and the spectra of the CH$_3$OH $J_K$=$7_0$--$6_0$ $A^+$ and N$_2$H$^+$ $J$=1--0 $F_1$=0--1 $F$=1--2 lines toward G034.43+00.24 MM1, MM2 and MM3. In the Spitzer image, G034.43+00.24 MM3, which is a MSX dark object, is less luminous than MM1 and MM2. It is known that a ultracompact HII region is associated with MM2 (Miralles et al. 1994; Shepherd et al. 2007). Thus, MM2 is thought to be most evolved among three objects, whereas MM3 is thought to be youngest.
In Figure \ref{fig:f9}, the CH$_3$OH spectrum of G034.43+00.24 MM3 is clearly broader than the N$_2$H$^+$ spectrum for G034.43+00.24 MM3. On the other hand, the CH$_3$OH velocity width is comparable to the N$_2$H$^+$ velocity width toward G034.43+00.24 MM1 and MM2. The CH$_3$OH linewidths toward these two sources are relatively narrower than that toward MM3.

As seen in Figure \ref{fig:f9}, the line shape of the CH$_3$OH emission sometimes differs from the Gaussian shape. It may consist of the broad wing component and the narrow component. In the present analysis, we fit all the CH$_3$OH spectra with a single Gaussian, so that the derived velocity widths have to be recognized as "effective" values for some sources where the line shape is much deviated from the single Gaussian shape. 
We employ the single Gaussian fitting rather than the double Gaussian fitting, because the latter fitting is difficult for most of the objects due to the limited S/N ratio.
Nevertheless, the "effective" line width reflects the presence of the broad component, and hence, our discussion on the basis of the line width would be justified at least qualitatively.

van der Tak et al. (2000a, 2000b) carried out multi-line observations of CH$_3$OH, C$^{34}$S, and C$^{17}$O toward several massive star-forming regions, and found that the velocity widths of the CH$_3$OH lines are broader than those of the C$^{34}$S and C$^{17}$O lines for the less luminous objects, whereas they are comparable in the luminous objects. van der Tak et al. (2000b) argued that shock by the interaction between an outflow and an ambient cloud is important as the origin of the gas-phase CH$_3$OH enhancement in low-luminosity objects, while radiation heating is important in high-luminosity objects.
In fact, the broad CH$_3$OH emission are reported for such an interacting region in several low-mass star forming regions, including the famous case of L1157, where a clear interaction between an outflow and an ambient cloud is observed as broad lines of CH$_3$OH, NH$_3$, and HC$_3$N (Umemoto et al. 1992; Mikami et al. 1992; Bachiller \& P\'{e}rez Guti\'{e}rrez 1997; Umemoto et al. 1999). 
In contrast, the N$_2$H$^+$ line is not detected in the interaction region.
Therefore, the CH$_3$OH $J_K$=$7_0$--$6_0$ $A^+$ emission would well trace the shocked region caused by the interaction between an outflow and an ambient cloud in the case of the MSX dark objects.
Furthermore, the broader line width can be seen in the observed lower-$J$ lines, $J_K$=$1_1$--$0_0$ $A^+$ and $4_0$--$4_{-1}$ $E$. This may imply that these CH$_3$OH emissions do not come from the quiescent and extended regions. To understand the properties of CH$_3$OH 
in the quiescent gas, we should observe the lower excitation transitions at the millimeter-wave region, since the critical density is proportional to $\nu^3$.
Our result that the line width of the CH$_3$OH line is broader in the MSX dark object means that the interaction between an outflow and an ambient cloud is important in the early stage of high-mass star formation.

\section{Summary}

\begin{itemize}

\item We have surveyed the N$_2$H$^+$ $J$=1--0, HC$_3$N $J$=5--4, CCS $J_N$=$4_3$--$3_2$, CH$_3$OH $J$=7--6, NH$_3$ (J, K) = (1, 1), (2, 2), and (3, 3) lines toward the massive clumps associated with the IRDCs. The N$_2$H$^+$ $J$=1--0 emission is detected toward 54 out of 55 objects, the HC$_3$N $J$=5--4 emission toward 43 objects, the NH$_3$ ($J$, $K$) = (1, 1) emission toward all the 55 objects, the NH$_3$ ($J$, $K$) = (2, 2) emission toward 52 objects, and the NH$_3$ ($J$, $K$) = (3, 3) emission toward 34 objects.
On the other hand, the CCS emission is not detected at all.
The CH$_3$OH $J$=7--6 emission is detected toward 18 out of 55 objects.
All the CH$_3$OH detected objects are associated with the Spitzer 24 $\mu$m sources.

\item The mean NH$_3$ rotation temperature of the MSX and Spitzer dark objects is 13.9 K, which tends to be lower than that of the objects with those of the MSX sources (17.9 K). 

\item The CCS column densities and the [CCS]/[N$_2$H$^+$] and [CCS]/[NH$_3$] ratios in the massive clumps associated with the IRDCs are lower than those in the low-mass starless cores. 
Thus, most of the massive clumps would be more chemically evolved than the low-mass starless cores. This survey demonstrates that the massive clump in the AFGL 333 region, which was found by Sakai et al. (2006, 2007), could be very novel.

\item The mean N$_2$H$^+$ velocity widths are significantly broader than those of the low-mass starless cores even in the MSX and Spitzer dark objects, indicating that massive cores associated with the IRDCs are initially turbulent.
This along with the chemical ages inferred from the [CCS]/[N$_2$H$^+$] ratio would be useful to constrain a models of star formation processes in IRDCs.

\item The velocity widths of the CH$_3$OH $J$=7--6 and NH$_3$ (3, 3) lines are broader than those of the other lines.  This seems to originate from the interaction between an outflow and an ambient cloud.
The CH$_3$OH velocity widths of the MSX dark objects are larger than those of the objects with MSX sources. This suggests that outflow interaction plays an important role in production of the gas phase CH$_3$OH in the early stage of high-mass star formation.  

\end{itemize}

\acknowledgments

We are grateful to the NRO staff for excellent support in the 45 m observations.
We would like to thank the members of the ASTE team for supporting the ASTE observations.
The 45 m radio telescope is operated by Nobeyama Radio Observatory, a branch of National Astronomical Observatory of Japan.
The ASTE project is driven by Nobeyama Radio Observatory, a branch of National Astronomical Observatory of Japan, in collaboration with University of Chile, and Japanese institutes including University of Tokyo, Nagoya University, Osaka Prefecture University, Ibaraki University, and Kobe University.




\clearpage


\begin{figure}
\figurenum{1}
 \epsscale{.90}
 \plotone{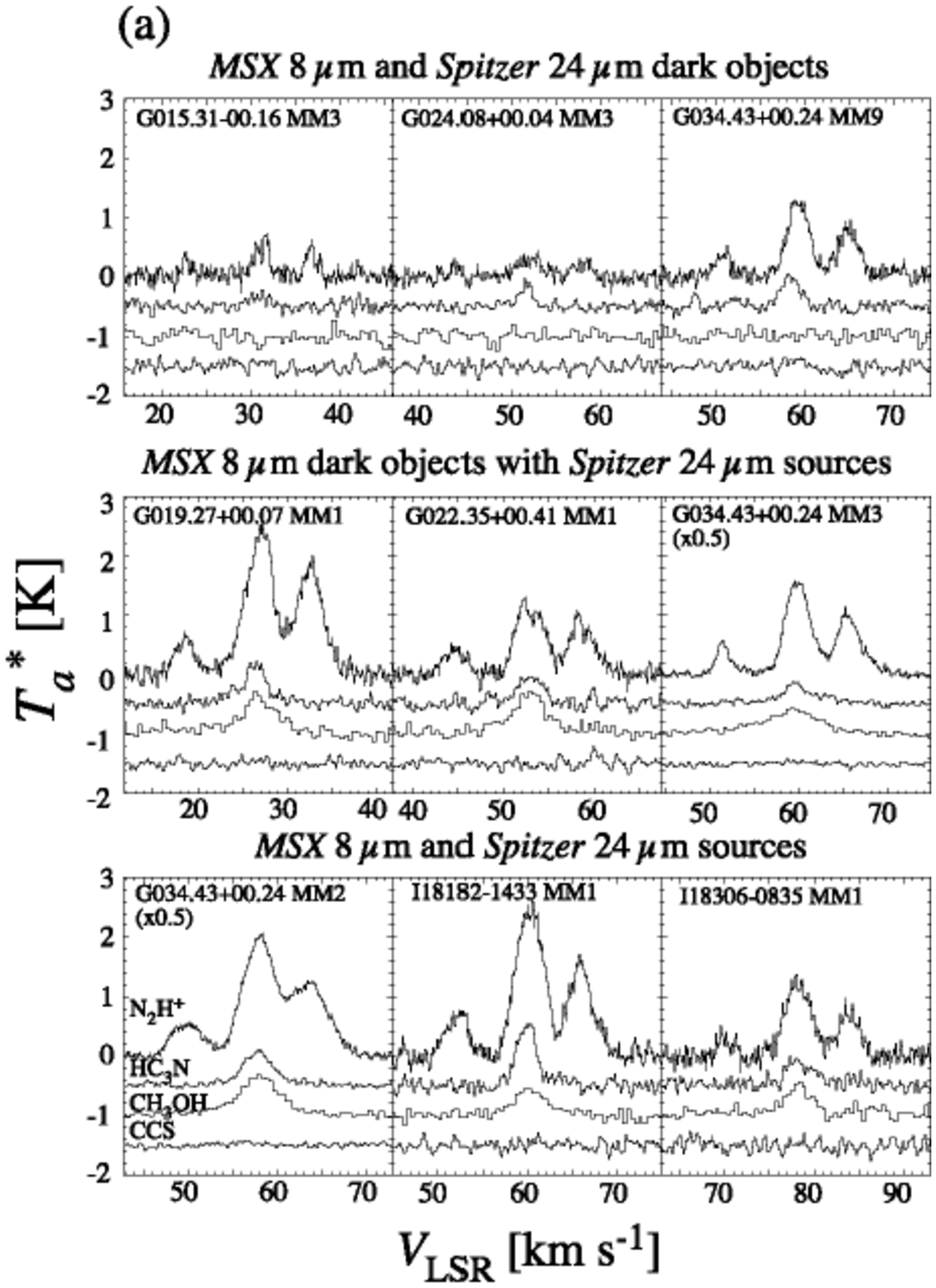}
	 \caption{Spectra of the N$_2$H$^+$ $J$=1--0, HC$_3$N $J$=5--4, CH$_3$OH $J_K$=$7_0$--$6_0$ $A^+$, and CCS $J_N$=$4_3$--$3_2$ lines toward the selected 9 objects (a) and all the other objects (b--e). 
For clarity, the spectra, except for N$_2$H$^+$, are offset from zero. In Figure 1a, the intensities for the spectra toward G034.43+00.24 MM2 and MM3 are divided by 2. Figures 1b--e are available in the electric edition of the journal. }
	\label{fig:f1}
\end{figure}
\clearpage

\begin{figure}
\figurenum{1b}
 \epsscale{.90}
 \plotone{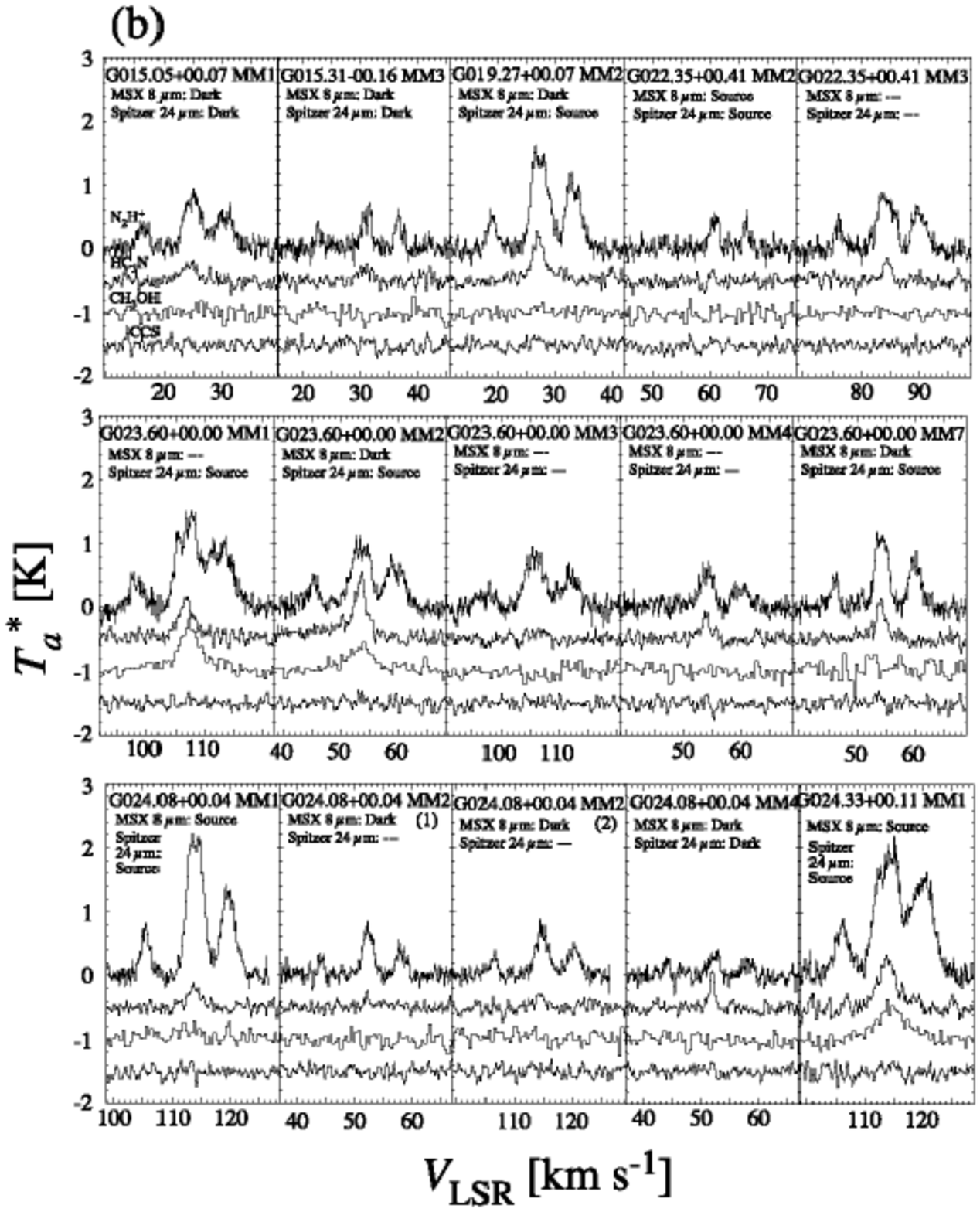}
	 \caption{Continued.}
	\label{fig:f1b}
\end{figure}

\clearpage

\begin{figure}
\figurenum{1c}
 \epsscale{.90}
     \plotone{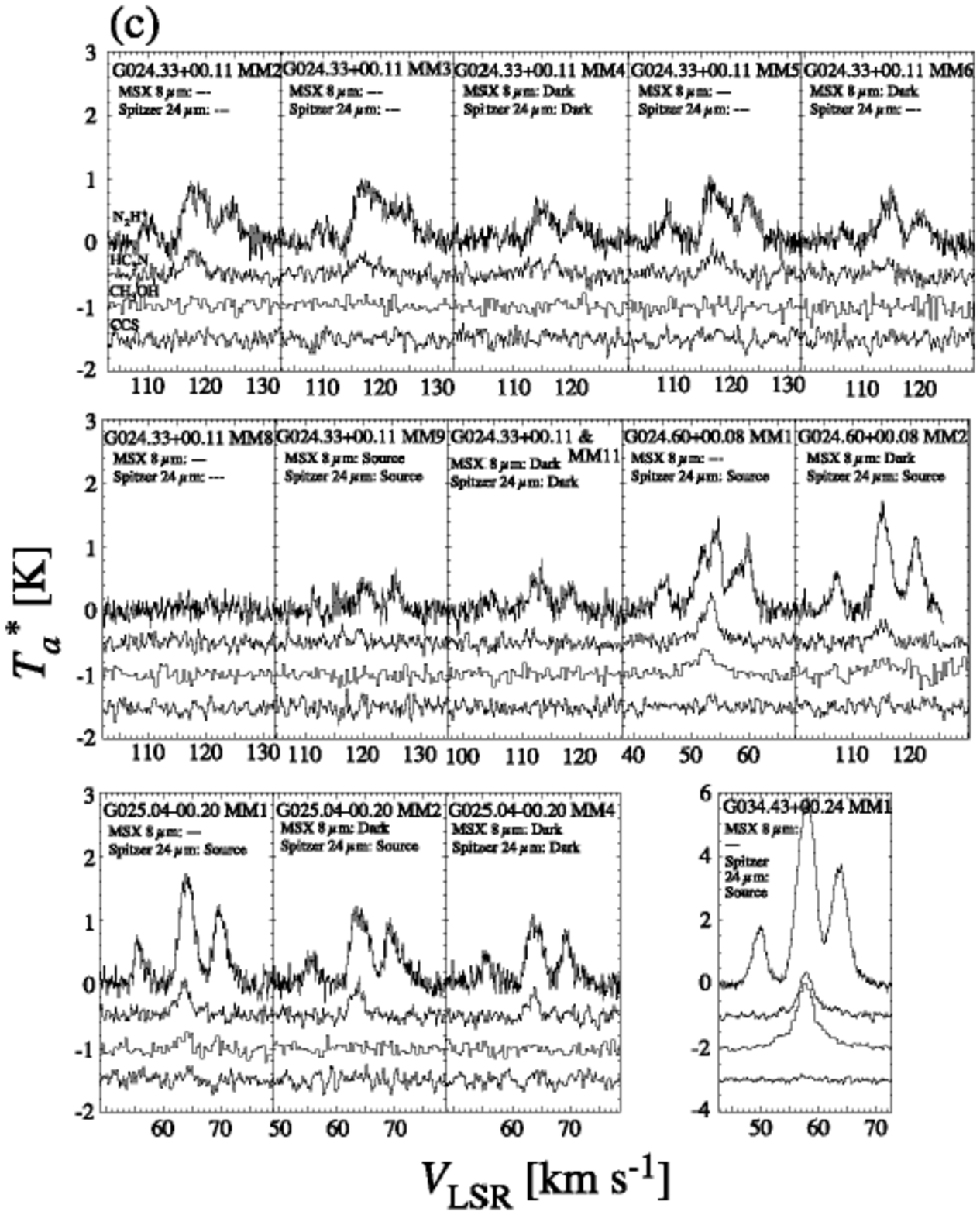}
	\caption{Continued.}
	\label{fig:f1c}
\end{figure}

\clearpage

\begin{figure}
\figurenum{1d}
 \epsscale{.90}
     \plotone{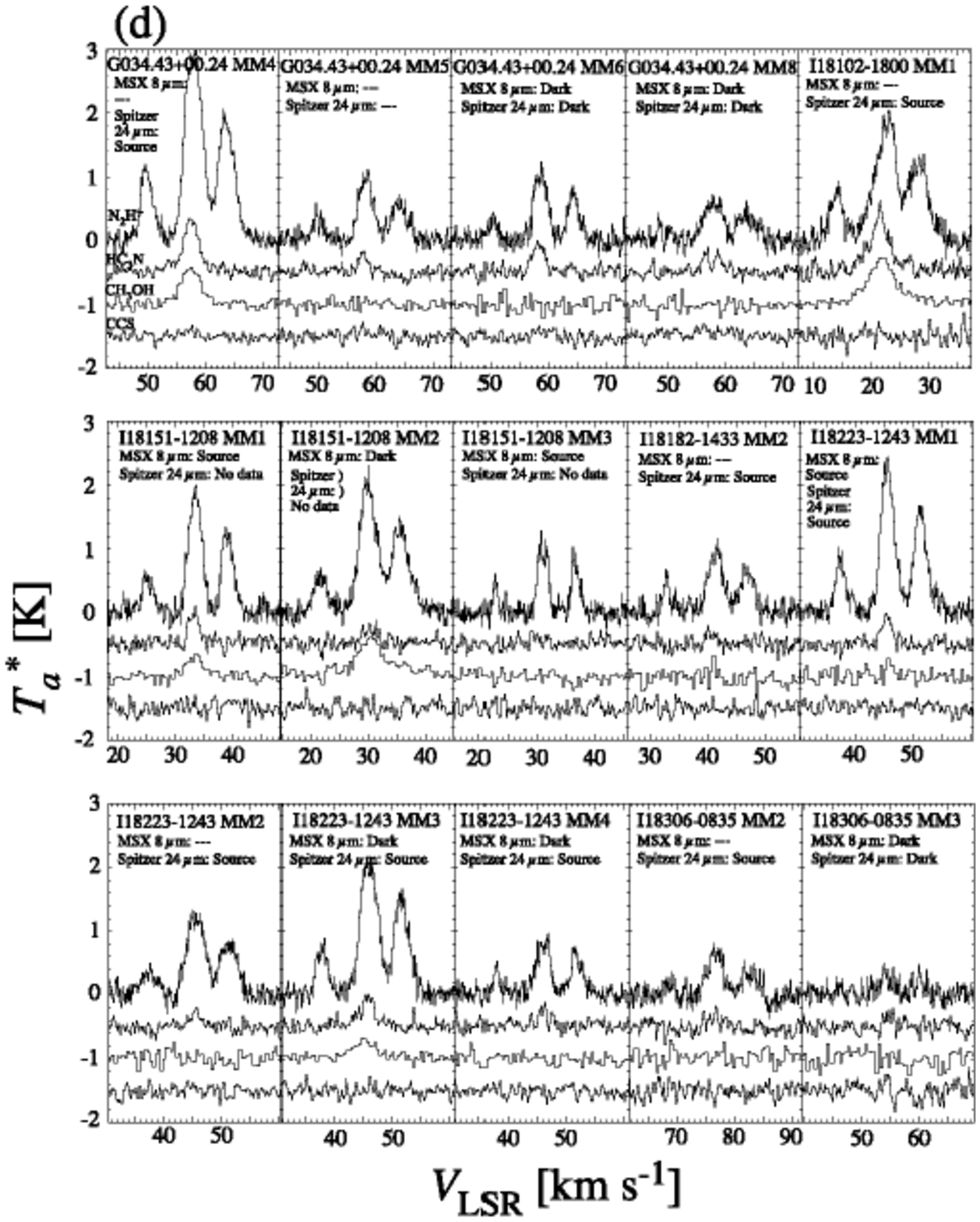}
	\caption{Continued.}
	\label{fig:f1d}
\end{figure}

\clearpage

\begin{figure}
\figurenum{1e}
 \epsscale{.60}
     \plotone{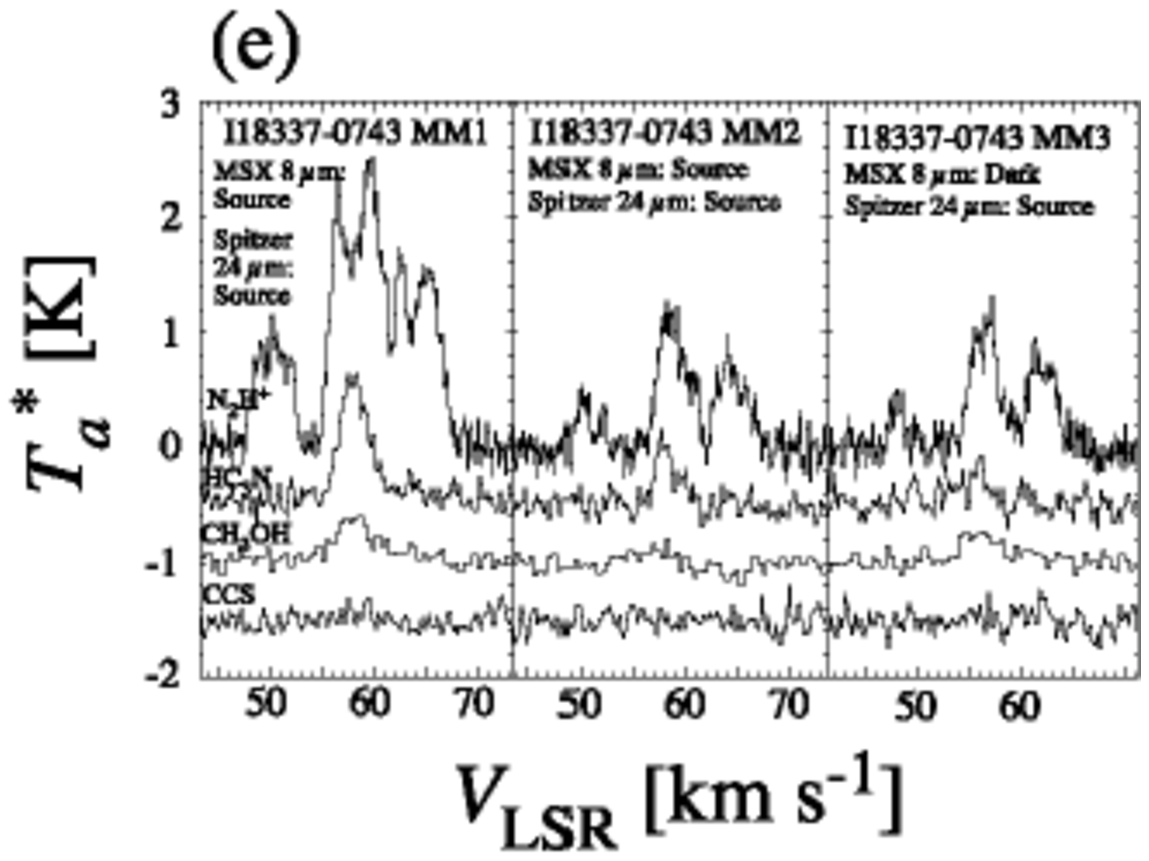}
	\caption{Continued.}
	\label{fig:f1e}
\end{figure}

\clearpage

\begin{figure}
\figurenum{2}
 \epsscale{.90}
 \plotone{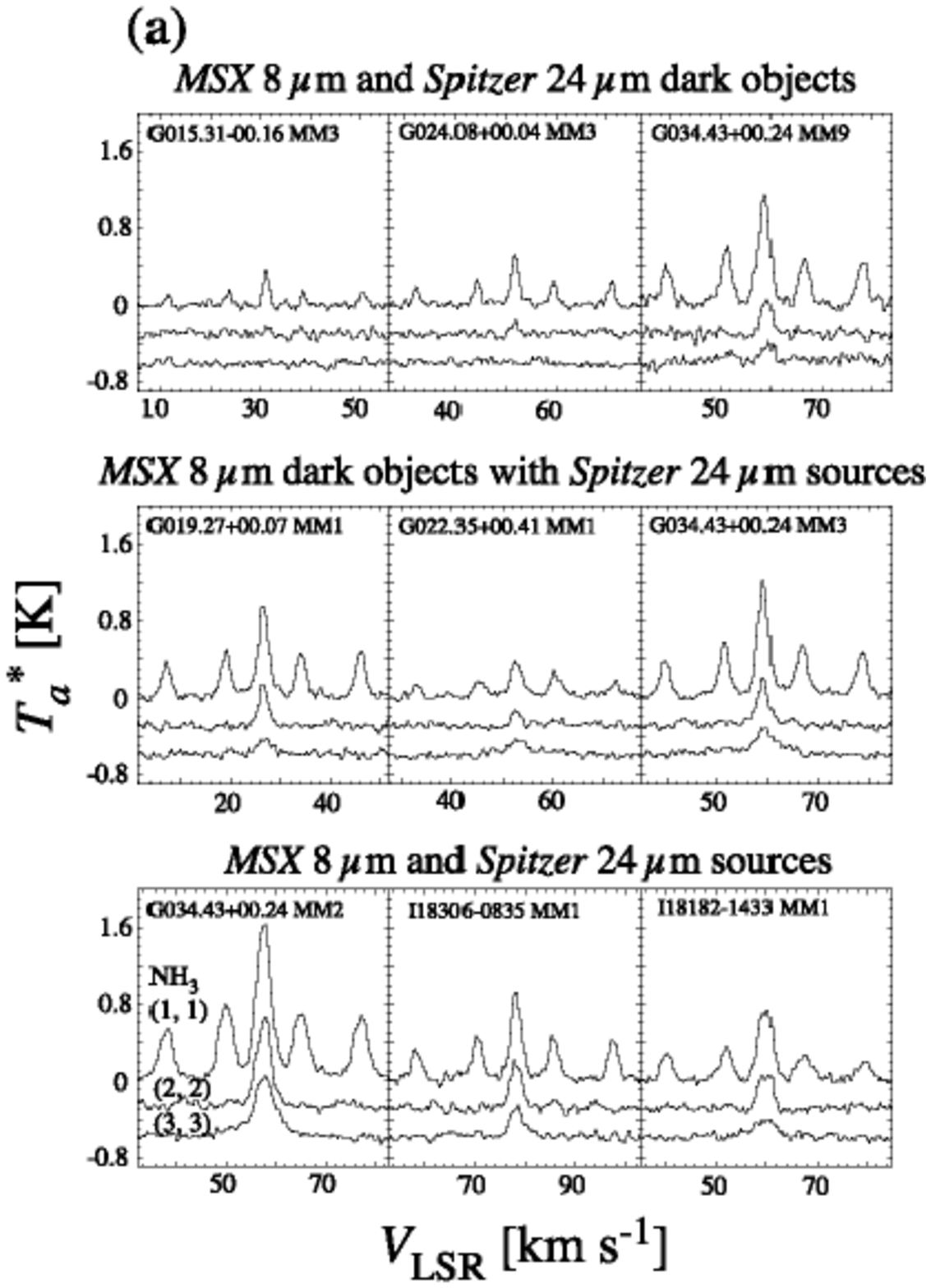}
	 \caption{Spectra of the NH$_3$ ($J$, $K$) = (1, 1), (2, 2), and (3, 3) lines toward the selected 9 objects (a) and all the other objects (b--e). For clarity, the spectra, except for the (1, 1) line, are offset from zero. Figures 2b--e are available in the electric edition of the journal.}
	\label{fig:f2}
\end{figure}

\clearpage

\begin{figure}
\figurenum{2b}
 \epsscale{.90}
 \plotone{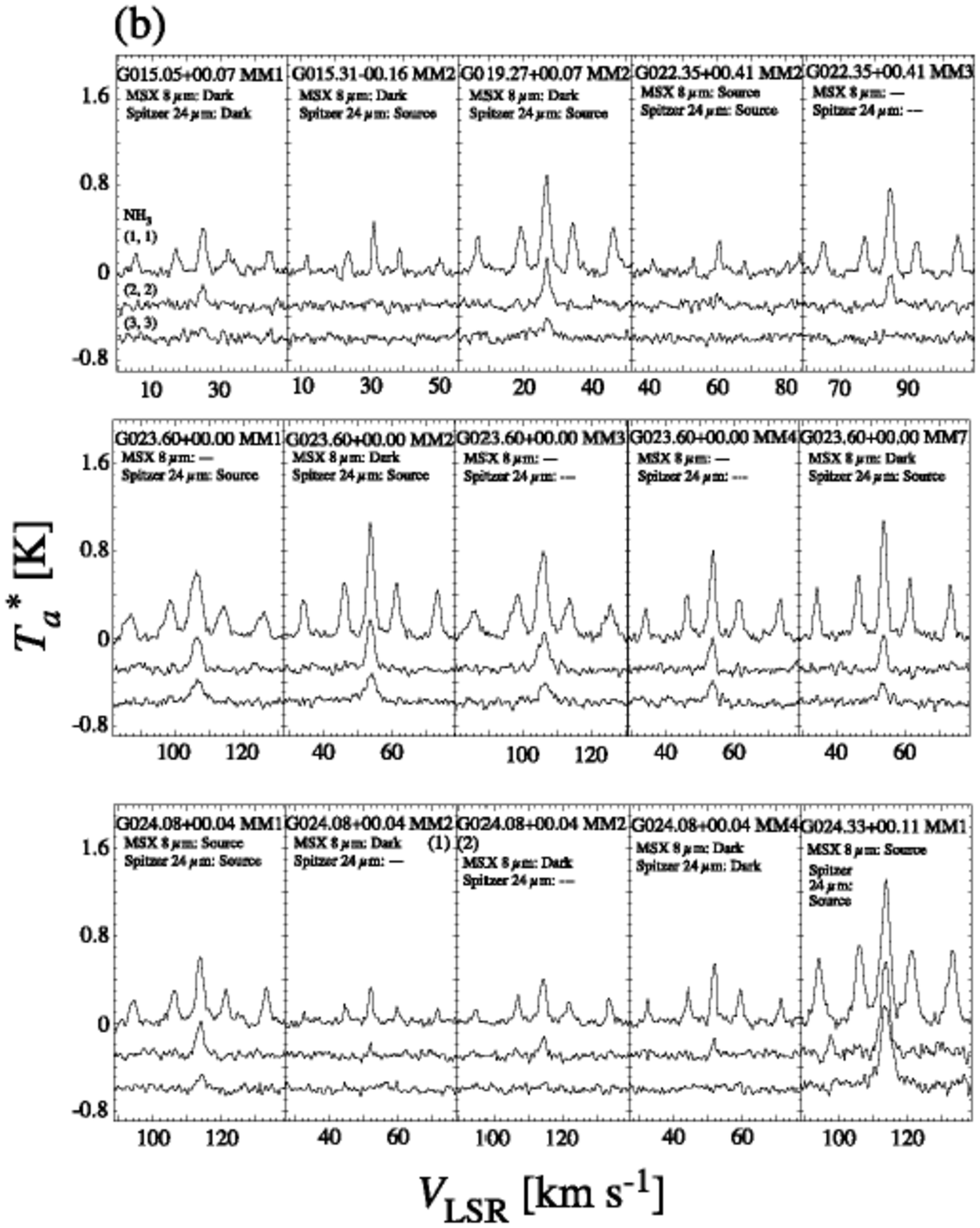}
	 \caption{Continued.}
	\label{fig:f2b}
\end{figure}

\clearpage

\begin{figure}
\figurenum{2c}
 \epsscale{.90}
     \plotone{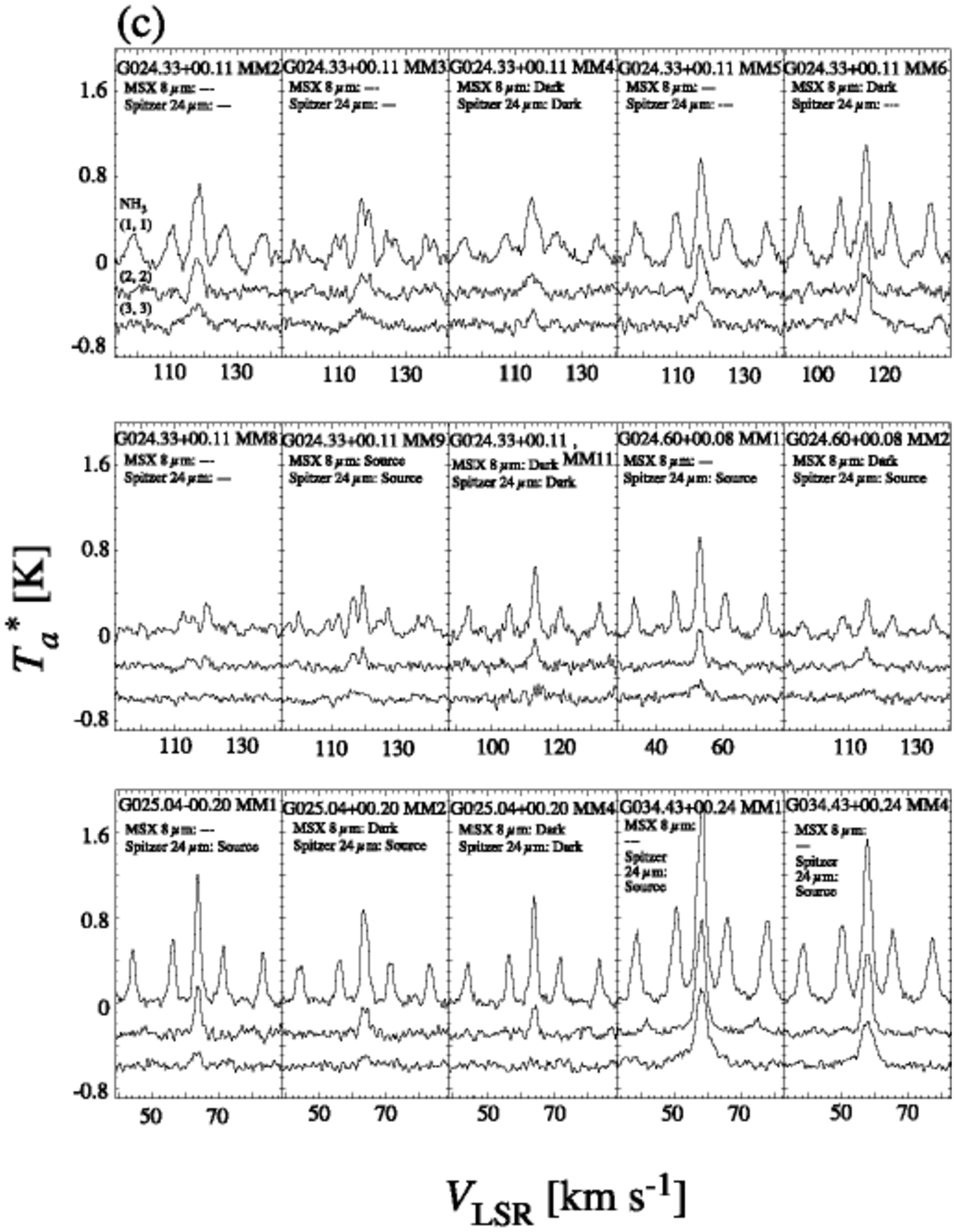}
	\caption{Continued.}
	\label{fig:f2c}
\end{figure}

\clearpage

\begin{figure}
\figurenum{2d}
 \epsscale{.90}
     \plotone{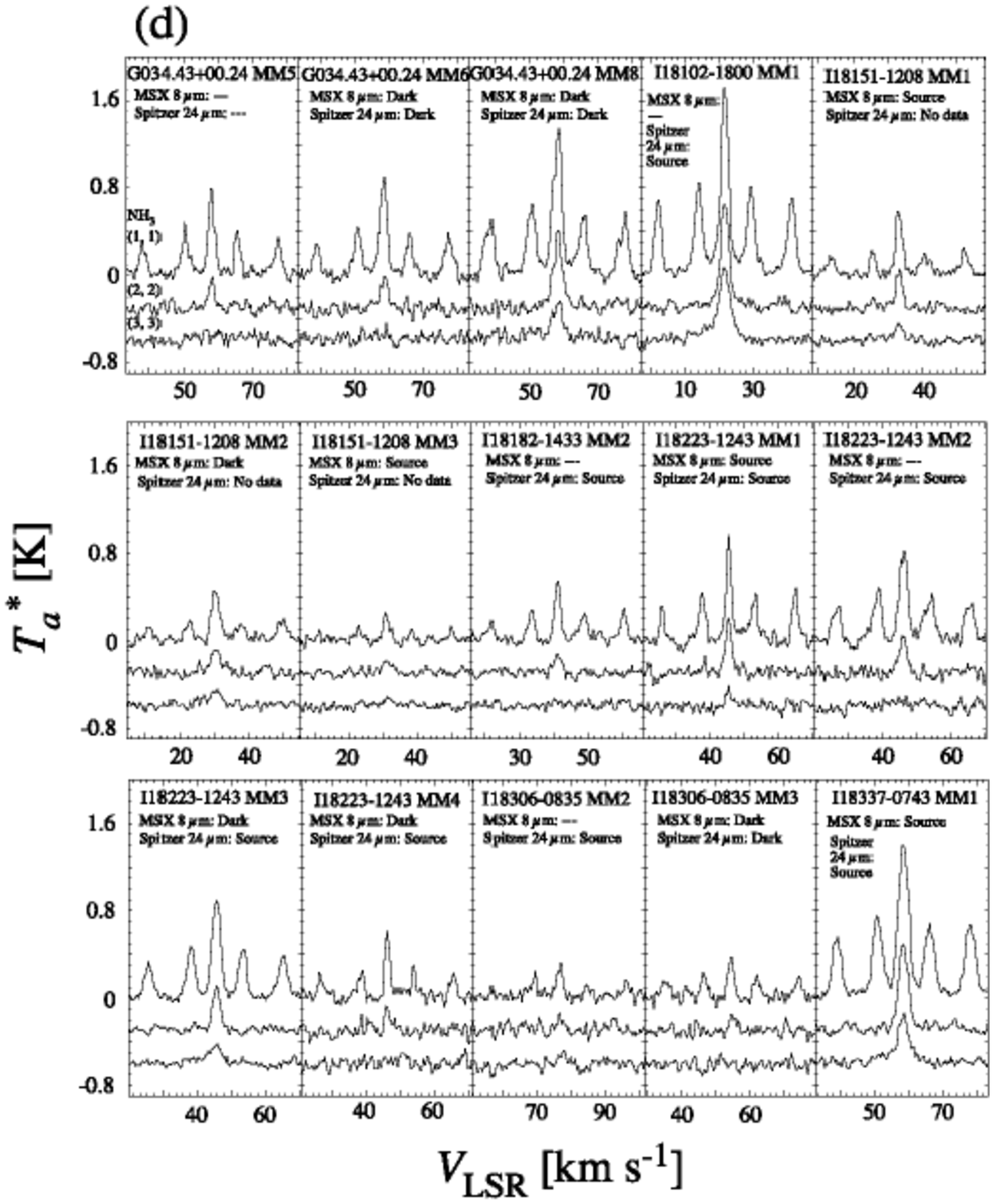}
	\caption{Continued.}
	\label{fig:f2d}
\end{figure}

\clearpage

\begin{figure}
\figurenum{2e}
 \epsscale{.50}
     \plotone{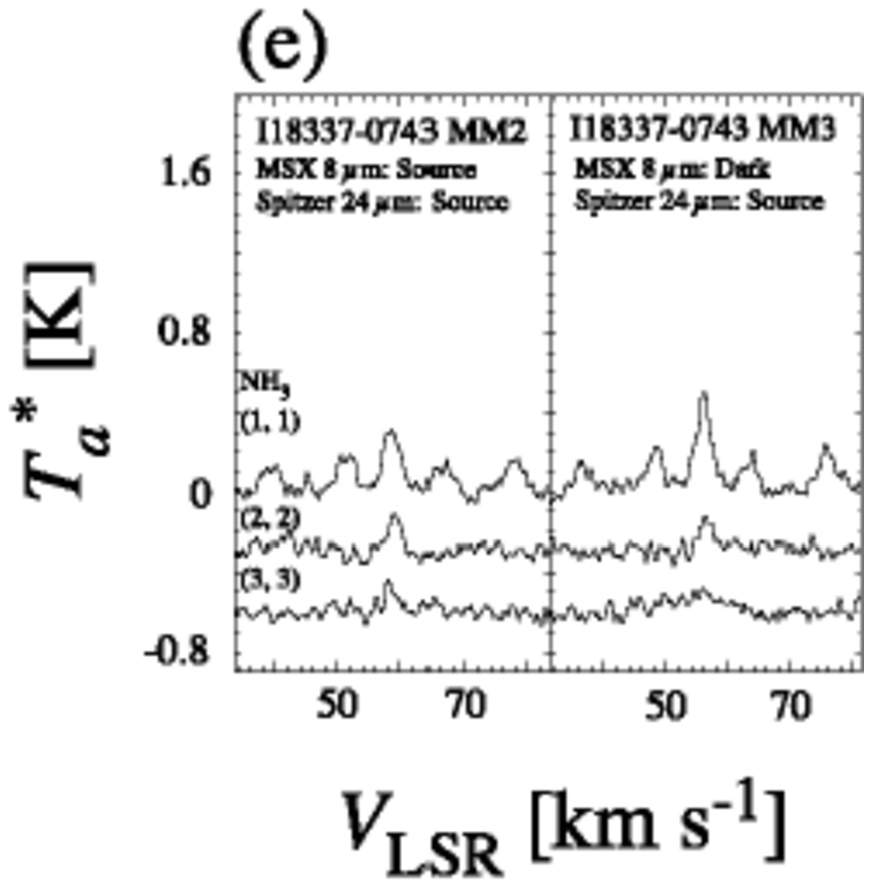}
	\caption{Continued.}
	\label{fig:f2e}
\end{figure}

\clearpage

\clearpage

\clearpage

\clearpage

\clearpage

\clearpage

\clearpage

\clearpage

\begin{deluxetable}{lrrrrrr}
\tablecolumns{7} 
\tablewidth{0pc} 
\tabletypesize{\small} 
\tablecaption{Target list.} 
\tablehead{ 
\colhead{Source} & \colhead{R. A.}   & \colhead{Dec.}    & \colhead{MSX\tablenotemark{a}}   & \colhead{Spitzer\tablenotemark{a}}   & \colhead{$F_{pk}$(1.2 mm)}&\colhead{Reference}\\
 & \footnotesize(J2000.0) & \footnotesize(J2000.0)  & 8 $\mu$m & 24 $\mu$m & [mJy] &
 }
\startdata 
G015.05+00.07 MM1 & 18 17 50.4 & -15 53 38 & D & D & 115 & 1\\
G015.31-00.16 MM2 & 18 18 50.4 & -15 43 19 & D & S & 39 & 1\\
G015.31-00.16 MM3 & 18 18 45.3 & -15 41 58 & D & D & 37 & 1\\
G019.27+00.07 MM1 & 18 25 58.5 & -12 03 59 & D & S & 150 & 1\\
G019.27+00.07 MM2 & 18 25 52.6 & -12 04 48 & D & S & 89 & 1\\
G022.35+00.41 MM1 & 18 30 24.4 & -09 10 34 & D & S & 349 & 1\\
G022.35+00.41 MM2 & 18 30 24.2 & -09 12 44 & S & S & 67 & 1\\
G022.35+00.41 MM3 & 18 30 38.1 & -09 12 44 & --- & --- & 53 & 1\\
G023.60+00.00 MM1 & 18 34 11.6 & -08 19 06 & --- & S & 375 & 1\\
G023.60+00.00 MM2 & 18 34 21.1 & -08 18 07 & D & S & 272 & 1\\
G023.60+00.00 MM3 & 18 34 10.0 & -08 18 28 & --- & --- & 81 & 1\\
G023.60+00.00 MM4 & 18 34 23.0 & -08 18 21 & --- & --- & 68 & 1\\
G023.60+00.00 MM7 & 18 34 21.1 & -08 17 11 & D & S & 58 & 1\\
G024.08+00.04 MM1 & 18 34 57.0 & -07 43 26 & S & S & 219 & 1\\
G024.08+00.04 MM2 & 18 34 51.1 & -07 45 32 & D & --- & 68 & 1\\
G024.08+00.04 MM3 & 18 35 02.2 & -07 45 25 & D & D & 50 & 1\\
G024.08+00.04 MM4 & 18 35 02.6 & -07 45 56 & D & D & 48 & 1\\
G024.33+00.11 MM1 & 18 35 07.9 & -07 35 04 & S & S & 1199 & 1\\
G024.33+00.11 MM2 & 18 35 34.5 & -07 37 28 & --- & --- & 117 & 1\\
G024.33+00.11 MM3 & 18 35 27.9 & -07 36 18 & --- & --- & 96 & 1\\
G024.33+00.11 MM4 & 18 35 19.4 & -07 37 17 & D & D & 90 & 1\\
G024.33+00.11 MM5 & 18 35 33.8 & -07 36 42 & --- & --- & 79 & 1\\
G024.33+00.11 MM6 & 18 35 07.7 & -07 34 33 & D & --- & 77 & 1\\
G024.33+00.11 MM8 & 18 35 23.4 & -07 37 21 & --- & --- & 72 & 1\\
G024.33+00.11 MM9 & 18 35 26.5 & -07 36 56 & S & S & 66 & 1\\
G024.33+00.11 MM11 & 18 35 05.1 & -07 35 58 & D & D & 48 & 1\\
G024.60+00.08 MM1 & 18 35 40.2 & -07 18 37 & --- & S & 279 & 1\\
G024.60+00.08 MM2 & 18 35 35.7 & -07 18 09 & D & S & 230 & 1\\
G025.04-00.20 MM1 & 18 38 10.2 & -07 02 34 & --- & S & 203 & 1\\
G025.04-00.20 MM2 & 18 38 17.7 & -07 02 51 & D & S & 92 & 1\\
G025.04-00.20 MM4 & 18 38 13.7 & -07 03 12 & D & D & 82 & 1\\
G034.43+00.24 MM1 & 18 53 18.0 & 01 25 24 & --- & S & 2228 & 1\\
G034.43+00.24 MM2 & 18 53 18.6 & 01 24 40 & S & S & 964 & 1\\
G034.43+00.24 MM3 & 18 53 20.4 & 01 28 23 & D & S & 244 & 1\\
G034.43+00.24 MM4 & 18 53 19.0 & 01 24 08 & --- & S & 221 & 1\\
G034.43+00.24 MM5 & 18 53 19.8 & 01 23 30 & --- & --- & 122 & 1\\
G034.43+00.24 MM6 & 18 53 18.6 & 01 27 48 & D & D & 57 & 1\\
G034.43+00.24 MM8 & 18 53 16.4 & 01 26 20 & D & D & 51 & 1\\
G034.43+00.24 MM9 & 18 53 18.4 & 01 28 14 & D & D & 50 & 1\\
I18102-1800 MM1 & 18 13 11.0 & -17 59 59 & --- & S & 316 & 2,3,4\\
I18151-1208 MM1 & 18 17 58.0 & -12 07 27 & S & No data & 672 & 2,3,4\\
I18151-1208 MM2 & 18 17 50.4 & -12 07 55 & D & No data & 424 & 2,3,4\\
I18151-1208 MM3 & 18 17 52.2 & -12 06 56 & S & No data & 149 & 2,3,4\\
I18182-1433 MM1 & 18 21 09.2 & -14 31 57 & S & S & 1303 & 2,3,4\\
I18182-1433 MM2 & 18 21 14.9 & -14 33 06 & --- & S & 100 & 2,3,4\\
I18223-1243 MM1 & 18 25 10.5 & -12 42 26 & S & S & 328 & 2,3,4\\
I18223-1243 MM2 & 18 25 09.5 & -12 44 15 & --- & S & 124 & 2,3,4\\
I18223-1243 MM3 & 18 25 08.3 & -12 45 28 & D & S & 205 & 2,3,4\\
I18223-1243 MM4 & 18 25 07.2 & -12 47 54 & D & S & 52 & 2,3,4\\
I18306-0835 MM1 & 18 33 24.0 & -08 33 31 & S & S & 731 & 2,3,4\\
I18306-0835 MM2 & 18 33 17.2 & -08 33 26 & --- & S & 212 & 2,3,4\\
I18306-0835 MM3 & 18 33 32.1 & -08 32 29 & D & D & 103 & 2,3,4\\
I18337-0743 MM1 & 18 36 41.0 & -07 39 20 & S & S & 485 & 2,3,4\\
I18337-0743 MM2 & 18 36 27.7 & -07 40 28 & S & S & 180 & 2,3,4\\
I18337-0743 MM3 & 18 36 18.2 & -07 41 01 & D & S & 110 & 2,3,4\\
\enddata 
\tablenotetext{a}{"D" indicates a dark object, and "S" indicates an object with point-like sources, whereas "---" represents an object without point-like sources and dark spots.}
\tablerefs{(1) Rathborne et al. (2006); (2) Sridharan et al. (2002); (3) Beuther et al. (2002); (4) Sridharan et al. (2005).}
\label{tab:t1}
\end{deluxetable} 

\clearpage

\begin{deluxetable}{llrrrrr}
\tablecolumns{7} 
\tablewidth{0pc} 
\tabletypesize{\small} 
\tablecaption{Observed lines.} 
\tablehead{ 
\colhead{Species} & \colhead{Transition}   & \colhead{$\nu$(rest)}   & \colhead{$\mu^2 S$}  & \colhead{$E_u$/$k$}    & \colhead{Reference} & \colhead{Telescope}\\
 & & [GHz]   & [Debye$^2$] & [K]&  & \\
 }
\startdata
CCS & $J_N$=$4_3$--$3_2$ & 45.379033 & 31.3 & 5.4 & 1 &  NRO 45 m \\
HC$_3$N & $J$=5--4 & 45.490302 & 69.2 & 6.5 & 2 & NRO 45 m \\
N$_2$H$^+$ & $J$=1--0 & 93.173777 & 11.6 &4.47 & 3 & NRO 45 m \\
NH$_3$ & ($J$, $K$) = (1, 1) & 23.694495 & 3.24 & 23.4& 4 & NRO 45 m \\
NH$_3$ & ($J$, $K$) = (2, 2) & 23.722633 & 7.20 & 64.9& 4 & NRO 45 m \\
NH$_3$ & ($J$, $K$) = (3, 3) & 23.870129 & 11.3 &124.5& 4 & NRO 45 m \\
CH$_3$OH & $J_K$ = $7_0$--$6_0$ $E$ & 338.124502 & 5.480 & 77.1 & 5 & ASTE \\
CH$_3$OH & $J_K$ = $7_{-1}$--$6_{-1}$ $E$ & 338.344628 & 5.375 & 69.6 & 5 & ASTE \\
CH$_3$OH & $J_K$ = $7_{0}$--$6_{0}$ $A^+$ & 338.408681 & 5.484 & 65.0 & 5 & ASTE \\
CH$_3$OH & $J_K$ = $7_{2}$--$6_{2}$ $A^-$ & 338.512762 & 5.059 & 86.4 & 5 & ASTE \\
CH$_3$OH & $J_K$ = $7_{3}$--$6_{3}$ $A^+$ & 338.540795 & 4.457 & 114.8 & 5 & ASTE \\
CH$_3$OH & $J_K$ = $7_{3}$--$6_{3}$ $A^-$ & 338.543204 & 4.457 & 114.8 & 5 & ASTE \\
CH$_3$OH & $J_K$ = $7_{-3}$--$6_{-3}$ $E$ & 338.559928 & 4.503 & 126.8 & 5 & ASTE \\
CH$_3$OH & $J_K$ = $7_{3}$--$6_{3}$ $E$ & 338.583195 & 4.486 & 111.8 & 5 & ASTE \\
CH$_3$OH & $J_K$ = $7_{1}$--$6_{1}$ $E$ & 338.614999 & 5.508 & 85.1 & 5 & ASTE \\
CH$_3$OH & $J_K$ = $7_{2}$--$6_{2}$ $A^+$ & 338.639939 & 5.509 & 102.7 & 5 & ASTE \\
CH$_3$OH & $J_K$ = $7_{2}$--$6_{2}$ $E$ & 338.721630 & 4.980 & 86.3 & 5 & ASTE \\
CH$_3$OH & $J_K$ = $7_{-2}$--$6_{-2}$ $E$ & 338.722940 & 5.046 & 90.0 & 5  & ASTE \\
CH$_3$OH & $J_K$ = $4_{0}$--$3_{-1}$ $E$ & 350.687730 & 1.553 & 35.4 & 5 & ASTE \\
CH$_3$OH & $J_K$ = $1_{1}$--$0_{0}$ $A^+$ & 350.905119 & 1.977 & 16.8 & 5 & ASTE \\
\enddata 
\tablerefs{(1) Yamamoto et al. (1990); (2) JPL Catalog (http://spec.jpl.nasa.gov/)  ; (3) Caselli et al. (1995) ; (4) Ho \& Townes (1983) ; (5) Anderson et al. (1990).}
\label{tab:t2}
\end{deluxetable} 

\clearpage

\begin{deluxetable}{llrrrrrrr} 
\tablecolumns{9} 
\tablewidth{0pc} 
\tablecaption{3 $\sigma$ detection rates of the observed lines.\tablenotemark{a}} 
\tablehead{ 
\colhead{MSX}& \colhead{Spitzer} & \colhead{N$_2$H$^+$}   & \colhead{HC$_3$N}    & \colhead{CCS}    & \colhead{CH$_3$OH} & \colhead{NH$_3$} &  \colhead{NH$_3$} &  \colhead{NH$_3$}     \\
& & $J$=1--0  & $J$=5--4   & $J_N$=$4_3$--$3_2$    &  $J$=7--6 &  (1, 1) &   (2, 2) &   (3, 3)    \\
 }
\startdata 
Dark&Dark& 11/11&7/11 & 0/11 & 0/11 &11/11 &10/11&3/11\\
Dark&Source&12/12&12/12& 0/12&6/12&12/12&11/12&8/12\\
Dark&Others&3/3&2/3&0/3&1/3&3/3&3/3&2/3\\
Source&Source&12/12&9/12&0/12&6/12&12/12&11/12&9/12\\
Others&Source&9/9&7/9&0/9&5/9&9/9&9/9&7/9\\
Others&Others&7/8&6/8&0/8&0/8&8/8&8/8&5/8\\ \hline
Total&&54/55&43/55&0/55&18/55&55/55&52/55&34/55\\
\enddata 
\tablenotetext{a}{Typical 3 $\sigma$ level in $T_a^*$ and channel width in velocity are 0.30 K and 0.070 km s$^{-1}$ for N$_2$H$^+$, 0.22 K and 0.134 km s$^{-1}$ for HC$_3$N, 0.24 K and 0.134 km s$^{-1}$ for CCS, 0.28 K and 0.443 km s$^{-1}$ for CH$_3$OH, and 0.08 K and 0.253 km s$^{-1}$ for NH$_3$.}
\label{tab:t3}
\end{deluxetable} 

\clearpage

\begin{deluxetable}{lrrrrrr} 
\tablecolumns{7} 
\tablewidth{0pc} 
\tabletypesize{\footnotesize} 
\tablecaption{Line parameters of the N$_2$H$^+$ $J$=1--0 line and distance of each object.\tablenotemark{a}} 
\tablehead{ 
\colhead{Source}& \colhead{$T_{\rm ex}$} &\colhead{$\tau$} & \colhead{$V_{LSR}$}   & \colhead{$\Delta V$} & \colhead{$\int T_a^* dV$ \tablenotemark{b}} & \colhead{Distance \tablenotemark{c}} \\
& [K] &  & [km s$^{-1}$] & [km s$^{-1}$] & [K km s$^{-1}$] &[kpc]\\
 }
\startdata 
G015.05+00.07 MM1 & 5.1 (0.2)& 2.5 (0.4)& 24.73 (0.03)& 2.79 (0.08)& 6.0 (0.1)& 2.5 \\
G015.31-00.16 MM2 & 4.0 (0.2)& 6.9 (2.0)& 31.12 (0.02)& 0.81 (0.07)& 2.3 (0.1)& 3.1 \\
G015.31-00.16 MM3 & 4.0 (0.1)& 11.4 (2.6)& 30.81 (0.02)& 0.62 (0.04)& 2.6 (0.1)& 3.0 \\
G019.27+0.07 MM1 & 15 (1)& 1.1 (0.1)& 26.76 (0.01)& 3.15 (0.03)& 17.7 (0.1)& 2.3 \\
G019.27+0.07 MM2 & 9.0 (0.6)& 1.5 (0.2)& 27.01 (0.02)& 2.62 (0.04)& 9.8 (0.1)& 2.3 \\
G022.35+00.41 MM1 & 5.9 (0.1)& 3.1 (0.3)& 52.73 (0.02)& 2.74 (0.05)& 8.9 (0.1)& 3.7 \\
G022.35+00.41 MM2 & 6 (2)& 1.50 (1.48)& 60.22 (0.02)& 0.83 (0.07)& 1.9 (0.1)& 4.1 \\
G022.35+00.41 MM3 & 5.0 (0.2)& 3.4 (0.5)& 83.97 (0.03)& 2.23 (0.07)& 5.4 (0.1)& 5.3 \\
G023.60+00.00 MM1 & 7.9 (0.5)& 1.5 (0.2)& 106.89 (0.06)& 3.82 (0.07)& 11.8 (0.1)& 6.7 \\
G023.60+00.00 MM2 & 5.8 (0.3)& 2.2 (0.3)& 53.34 (0.03)& 3.03 (0.07)& 7.6 (0.1)& 3.7 \\
G023.60+00.00 MM3 & 8 (3)& 0.7 (0.4)& 105.51 (0.04)& 3.4 (0.1)& 5.8 (0.1)& 6.6 \\
G023.60+00.00 MM4 & 13 (24)& 0.2 (0.6)& 53.87 (0.05)& 2.6 (0.1)& 3.0 (0.1)& 3.7 \\
G023.60+00.00 MM7 & 8 (1)& 1.3 (0.3)& 54.00 (0.05)& 1.90 (0.06)& 6.0 (0.1)& 3.7 \\
G024.08+00.04 MM1 & 49 (26)& 0.21 (0.13)& 113.63 (0.01)& 2.67 (0.03)& 12.8 (0.1)& 7.8 \\
G024.08+00.04 MM2(1) & 9.4 (8.5)& 0.6 (0.9)& 52.08 (0.03)& 1.5 (0.1)& 2.8 (0.1)& 3.6 \\
G024.08+00.04 MM2(2) & 7 (3)& 0.8 (0.7)& 114.45 (0.04)& 2.1 (0.1)& 3.6 (0.1)& 7.8 \\
G024.08+00.04 MM3 & 3.5 (0.1)& 3.5 (1)& 51.86 (0.06)& 2.2 (0.2)& 2.0 (0.1)& 3.6 \\
G024.08+00.04 MM4 & 3.32 (0.04)& 19 (6)& 51.95 (0.03)& 0.85 (0.08)& 1.7 (0.1)& 3.6 \\
G024.33+00.11 MM1 & 7.6 (0.1)& 3.4 (0.2)& 113.76 (0.02)& 3.53 (0.04)& 19.0 (0.1)& 7.7 \\
G024.33+00.11 MM2 & 5.4 (0.3)& 2.3 (0.4)& 118.22 (0.04)& 3.23 (0.09)& 6.8 (0.1)& 7.7 \\
G024.33+00.11 MM3 & 8 (2)& 0.8 (0.3)& 117.78 (0.05)& 4.3 (0.1)& 7.9 (0.1)& 7.7 \\
G024.33+00.11 MM4 & 6 (2)& 0.9 (0.6)& 115.19 (0.07)& 3.6 (0.2)& 4.2 (0.1)& 7.7 \\
G024.33+00.11 MM5 & 4.69 (0.09)& 4.6 (0.5)& 117.26 (0.04)& 3.00 (0.08)& 7.6 (0.1)& 7.7 \\
G024.33+00.11 MM6 & 7 (2)& 0.9 (0.5)& 114.43 (0.04)& 2.97 (0.10)& 4.5 (0.1)& 7.7 \\
G024.33+00.11 MM8 & --- & --- & --- & --- & 0.9 (0.1)& 7.7 \\
G024.33+00.11 MM9 & 3.8 (0.2)& 4.2 (1.5)& 119.78 (0.05)& 1.7 (0.2)& 2.7 (0.1)& 7.7 \\
G024.33+00.11 MM11 & 4.3 (0.3)& 2.4 (0.9)& 112.60 (0.05)& 2.18 (0.05)& 3.0 (0.1)& 7.7 \\
G024.60+00.08 MM1 & 5.4 (0.1)& 3.9 (0.3)& 53.14 (0.03)& 3.10 (0.06)& 9.8 (0.1)& 3.6 \\
G024.60+00.08 MM2 & 10 (2)& 1.2 (0.2)& 115.16 (0.01)& 2.37 (0.04)& 9.3 (0.1)& 7.7 \\
G025.04-00.20 MM1 & 11 (1)& 1.1 (0.2)& 64.00 (0.02)& 2.36 (0.05)& 9.8 (0.1)& 4.2 \\
G025.04-00.20 MM2 & 6.5 (0.4)& 2.0 (0.3)& 63.83 (0.03)& 2.77 (0.07)& 8.1 (0.1)& 4.2 \\
G025.04-00.20 MM4 & 6.2 (0.5)& 1.9 (0.4)& 63.69 (0.02)& 2.18 (0.08)& 6.0 (0.1)& 4.1 \\
G034.43+00.24 MM1 & 39 (2)& 0.75 (0.05)& 57.77 (0.05)& 3.15 (0.01)& 40.5 (0.1)& 3.5 \\
G034.43+00.24 MM2 & 29 (2)& 0.67 (0.06)& 57.832 (0.009)& 4.04 (0.02)& 34.2 (0.1)& 3.5 \\
G034.43+00.24 MM3 & 36 (7)& 0.42 (0.09)& 59.686 (0.008)& 2.90 (0.02)& 20.3 (0.1)& 3.6 \\
G034.43+00.24 MM4 & 14.2 (0.5)& 1.5 (0.1)& 57.689 (0.008)& 2.81 (0.02)& 20.8 (0.1)& 3.5 \\
G034.43+00.24 MM5 & 9 (2)& 0.9 (0.3)& 58.12 (0.03)& 2.58 (0.08)& 6.0 (0.1)& 3.5 \\
G034.43+00.24 MM6 & 10 (2)& 0.8 (0.4)& 58.49 (0.02)& 2.07 (0.07)& 5.6 (0.1)& 3.6 \\
G034.43+00.24 MM8 & 6 (1)& 1.1 (0.5)& 57.63 (0.05)& 3.2 (0.1)& 4.5 (0.1)& 3.5 \\
G034.43+00.24 MM9 & 12 (3)& 0.7 (0.3)& 59.00 (0.02)& 2.45 (0.06)& 7.3 (0.1)& 3.6 \\
I18102-1800 MM1 & 9.3 (0.5)& 1.7 (0.2)& 22.36 (0.02)& 3.49 (0.05)& 15.9 (0.1)& 2.7 \\
I18151-1208 MM1 & 13 (2)& 0.8 (0.2)& 33.19 (0.01)& 2.10 (0.04)& 9.8 (0.1)& 2.8 \\
I18151-1208 MM2 & 14 (2)& 0.9 (0.2)& 29.74 (0.02)& 3.09 (0.04)& 14.2 (0.1)& 2.6 \\
I18151-1208 MM3 & 5.7 (0.2)& 4.9 (0.9)& 30.65 (0.01)& 0.87 (0.04)& 4.4 (0.1)& 2.7 \\
I18182-1433 MM1 & 30 (8)& 0.4 (0.1)& 60.00 (0.01)& 3.06 (0.04)& 16.6 (0.1)& 4.6 \\
I18182-1433 MM2 & 5.8 (0.4)& 2.3 (0.4)& 41.10 (0.03)& 2.28 (0.08)& 6.2 (0.1)& 3.6 \\
I18223-1243 MM1 & 14 (1)& 1.2 (0.2)& 45.32 (0.01)& 2.16 (0.03)& 13.4 (0.1)& 3.6 \\
I18223-1243 MM2 & 8.1 (0.7)& 1.3 (0.2)& 45.29 (0.02)& 2.99 (0.06)& 8.7 (0.1)& 3.6 \\
I18223-1243 MM3 & 9.4 (0.3)& 2.2 (0.2)& 45.73 (0.01)& 2.64 (0.03)& 14.5 (0.1)& 3.7 \\
I18223-1243 MM4 & 5.3 (0.3)& 2.4 (0.5)& 45.83 (0.03)& 1.97 (0.07)& 5.0 (0.1)& 3.7 \\
I18306-0835 MM1 & 27 (31)& 0.2 (0.3)& 78.02 (0.03)& 2.81 (0.07)& 7.2 (0.1)& 4.9 \\
I18306-0835 MM2 & 10 (11)& 0.4 (0.6)& 76.67 (0.05)& 2.6 (0.1)& 3.4 (0.1)& 4.9 \\
I18306-0835 MM3 & 5 (2)& 0.95 (1.5)& 53.98 (0.09)& 2.1 (0.3)& 1.6 (0.1)& 3.7 \\
I18337-0743 MM1 & 7.03 (0.08)& 5.9 (0.3)& 58.37 (0.02)& 3.35 (0.04)& 21.5 (0.1)& 3.9 \\
I18337-0743 MM2 & 8 (1)& 1.3 (0.4)& 58.60 (0.03)& 2.71 (0.08)& 6.9 (0.1)& 3.9 \\
I18337-0743 MM3 & 6.7 (0.6)& 1.7 (0.4)& 56.37 (0.03)& 2.44 (0.08)& 6.6 (0.1)& 3.8 \\
\enddata 
\tablenotetext{a}{The quoted errors denote one standard deviation.}
\tablenotetext{b}{Velocity range for integration is $\pm10$ km s$^{-1}$ centered at the $LSR$ velocity of N$_2$H$^+$ $J$=1--0 or NH$_3$ (1, 1).}
\tablenotetext{c}{The distance is the assumed kinematic distance based on the rotation curve of Clemens (1985).}
\label{tab:t4}
\end{deluxetable} 

\clearpage

\begin{deluxetable}{lrrrrrrr} 
\tablecolumns{8} 
\tablewidth{0pc} 
\tabletypesize{\footnotesize} 
\tablecaption{Line parameters of the HC$_3$N $J$=5--4 and CCS $J_N$=$4_3$--$3_2$ lines.\tablenotemark{a}} 
\tablehead{ 
&\multicolumn{4}{c}{HC$_3$N} & &\multicolumn{2}{c}{CCS} \\ \cline{2-5}\cline{7-8}
\colhead{Source}& \colhead{$T_{\rm pk}$} & \colhead{$V_{LSR}$}   & \colhead{$\Delta V$} & \colhead{$\int T_a^* dV$\tablenotemark{b}} & &\colhead{$T_{\rm pk}$} & \colhead{$\int T_a^* dV$\tablenotemark{b}} \\
& [K] & [km s$^{-1}$] & [km s$^{-1}$] & [K km s$^{-1}$] & &[K] & [K km s$^{-1}$]\\
 }
\startdata 
G015.05+00.07 MM1 & 0.25 (0.02)& 24.4 (0.1)& 3.5 (0.3)& 1.05 (0.09)& &$<$0.22  & $<$0.25 \\
G015.31-00.16 MM2 & 0.28 (0.03)& 30.92 (0.08)& 1.4 (0.2)& 0.61 (0.09)&& $<$0.24  & $<$0.28 \\
G015.31-00.16 MM3 & $<$0.23 & ---&  --- & 0.51 (0.09)&& $<$0.25  & $<$0.29 \\
G019.27+0.07 MM1 & 0.60 (0.02)& 26.39 (0.05)& 3.7 (0.1)& 2.60 (0.07)& &$<$0.19  & $<$0.23 \\
G019.27+0.07 MM2 & 0.64 (0.02)& 27.02 (0.04)& 2.9 (0.1)& 2.35 (0.07)& &$<$0.20  & $<$0.23 \\
G022.35+00.41 MM1 & 0.46 (0.02)& 53.00 (0.07)& 3.6 (0.2)& 1.94 (0.08)& &$<$0.20  & $<$0.23 \\
G022.35+00.41 MM2 & $<$0.20 &  --- &   --- & $<$0.24 & &$<$0.21  & $<$0.24 \\
G022.35+00.41 MM3 & 0.34 (0.03)& 84.36 (0.06)& 1.6 (0.1)& 0.59 (0.07)& &$<$0.19  & $<$0.22 \\
G023.60+00.00 MM1 & 0.57 (0.02)& 106.7 (0.06)& 3.6 (0.1)& 2.28 (0.08)&& $<$0.23  & $<$0.26 \\
G023.60+00.00 MM2 & 0.81 (0.02)& 53.37 (0.04)& 3.4 (0.1)& 3.35 (0.08)& &$<$0.20  & $<$0.23 \\
G023.60+00.00 MM3 & $<$0.21 &  --- &  --- & 0.41 (0.08)&& $<$0.23  & $<$0.27 \\
G023.60+00.00 MM4 & 0.33 (0.02)& 53.91 (0.07)& 2.3 (0.2)& 0.76 (0.07)& &$<$0.21  & $<$0.25 \\
G023.60+00.00 MM7 & 0.55 (0.02)& 54.08 (0.05)& 2.1 (0.1)& 1.56 (0.08)& &$<$0.21  & $<$0.24 \\
G024.08+00.04 MM1 & 0.30 (0.02)& 113.68 (0.08)& 2.8 (0.2)& 0.89 (0.07)& &$<$0.21  & $<$0.24 \\
G024.08+00.04 MM2(1) & 0.22 (0.03)& 52.18 (0.07)& 0.9 (0.2)& 0.44 (0.07)& &$<$0.21  & $<$0.24 \\
G024.08+00.04 MM2(2) & $<$0.19 & --- &  --- & 0.38 (0.07)&& $<$0.21  & $<$0.24 \\
G024.08+00.04 MM3 & 0.39 (0.02)& 51.67 (0.04)& 1.6 (0.1)& 0.78 (0.07)&& $<$0.24  & $<$0.28 \\
G024.08+00.04 MM4 & 0.56 (0.03)& 51.90 (0.02)& 1.08 (0.07)& 0.83 (0.07)&& $<$0.21  & $<$0.24 \\
G024.33+00.11 MM1 & 0.71 (0.02)& 113.69 (0.06)& 4.0 (0.1)& 3.22 (0.09)&& $<$0.28  & $<$0.32 \\
G024.33+00.11 MM2 & 0.34 (0.02)& 118.1 (0.1)& 3.4 (0.2)& 1.27 (0.09)&& $<$0.25  & $<$0.29 \\
G024.33+00.11 MM3 & 0.24 (0.02)& 117.2 (0.2)& 5.1 (0.4)& 1.31 (0.09)&& $<$0.27  & $<$0.31 \\
G024.33+00.11 MM4 & $<$0.23 & --- & --- & 1.11 (0.09)&& $<$0.26  & $<$0.30 \\
G024.33+00.11 MM5 & 0.34 (0.02)& 117.4 (0.1)& 4.0 (0.3)& 1.48 (0.09)&& $<$0.23  & $<$0.27 \\
G024.33+00.11 MM6 & $<$0.24 & --- & --- & 0.73 (0.09)&& $<$0.28  & $<$0.32 \\
G024.33+00.11 MM8 & $<$0.23 & --- & --- & $<$0.27&& $<$0.29  & $<$0.34 \\
G024.33+00.11 MM9 & 0.21 (0.03)& 119.4 (0.1)& 1.3 (0.2)& 0.50 (0.09)&& $<$0.29  & $<$0.33 \\
G024.33+00.11 MM11 & $<$0.23 & --- & --- & $<$0.27&& $<$0.28  & $<$0.32 \\
G024.60+00.08 MM1 & 0.63 (0.02)& 53.22 (0.05)& 3.5 (0.1)& 2.59 (0.08)&& $<$0.17  & $<$0.20 \\
G024.60+00.08 MM2 & 0.29 (0.02)& 115.2 (0.1)& 3.0 (0.2)& 1.23 (0.09)&& $<$0.28  & $<$0.32 \\
G025.04-00.20 MM1 & 0.46 (0.02)& 63.67 (0.07)& 3.0 (0.2)& 1.67 (0.09)&& $<$0.23  & $<$0.26 \\
G025.04-00.20 MM2 & 0.46 (0.03)& 63.63 (0.07)& 2.4 (0.2)& 1.21 (0.09)&& $<$0.29  & $<$0.34 \\
G025.04-00.20 MM4 & 0.35 (0.03)& 63.76 (0.08)& 2.0 (0.2)& 0.67 (0.09)&& $<$0.23  & $<$0.27 \\
G034.43+00.24 MM1 & 1.22 (0.02)& 58.04 (0.03)& 3.85 (0.06)& 5.35 (0.06)&& $<$0.18  & 0.35 (0.07)\\
G034.43+00.24 MM2 & 1.10 (0.01)& 57.86 (0.03)& 4.22 (0.06)& 5.09 (0.06)&& $<$0.18  & 0.43 (0.07)\\
G034.43+00.24 MM3 & 0.57 (0.01)& 60.05 (0.07)& 5.8 (0.2)& 3.29 (0.07)&& $<$0.18  & 0.41 (0.07)\\
G034.43+00.24 MM4 & 0.80 (0.02)& 57.50 (0.03)& 3.22 (0.08)& 3.03 (0.07)&& $<$0.17  & 0.30 (0.07)\\
G034.43+00.24 MM5 & 0.31 (0.03)& 57.81 (0.08)& 1.8 (0.2)& 0.67 (0.08)&& $<$0.22  & 0.38 (0.08)\\
G034.43+00.24 MM6 & 0.45 (0.02)& 58.34 (0.06)& 2.5 (0.1)& 1.25 (0.08)&& $<$0.23  & $<$0.26 \\
G034.43+00.24 MM8 & 0.23 (0.02)& 57.7 (0.2)& 4.4 (0.4)& 1.1 (0.08)&& $<$0.21  & $<$0.24 \\
G034.43+00.24 MM9 & 0.50 (0.02)& 58.40 (0.05)& 2.4 (0.1)& 1.31 (0.08)&& $<$0.21  & $<$0.24 \\
I18102-1800 MM1 & 0.83 (0.02)& 21.53 (0.05)& 4.4 (0.1)& 4.14 (0.09)&& $<$0.29  & $<$0.34 \\
I18151-1208 MM1 & 0.53 (0.03)& 33.26 (0.05)& 2.1 (0.1)& 1.39 (0.09)&& $<$0.26  & $<$0.3 \\
I18151-1208 MM2 & 0.25 (0.02)& 30.3 (0.1)& 3.2 (0.3)& 0.93 (0.10)&& $<$0.27  & $<$0.32 \\
I18151-1208 MM3 & $<$0.24 & --- & --- & 0.39 (0.09)&& $<$0.28  & $<$0.32 \\
I18182-1433 MM1 & 1.04 (0.03)& 59.86 (0.03)& 2.58 (0.07)& 3.07 (0.09)&& $<$0.26  & $<$0.3 \\
I18182-1433 MM2 & $<$0.25 & --- &  --- & 0.39 (0.10)&& $<$0.26  & $<$0.3 \\
I18223-1243 MM1 & 0.44 (0.03)& 45.38 (0.06)& 2.1 (0.1)& 1.09 (0.08)&& $<$0.22  & $<$0.25 \\
I18223-1243 MM2 & 0.24 (0.03)& 45.5 (0.1)& 2.1 (0.3)& 0.68 (0.09)& &$<$0.21  & $<$0.25 \\
I18223-1243 MM3 & 0.44 (0.02)& 45.65 (0.07)& 3.2 (0.2)& 1.59 (0.08)&& $<$0.24  & $<$0.28 \\
I18223-1243 MM4 & $<$0.22 & --- &  --- & 0.71 (0.08)& &$<$0.22  & $<$0.25 \\
I18306-0835 MM1 & 0.39 (0.02)& 78.5 (0.1)& 4.0 (0.2)& 1.7 (0.09)& &$<$0.30  & $<$0.34 \\
I18306-0835 MM2 & $<$0.25 & --- &  --- & 0.41 (0.10)&& $<$0.31  & $<$0.36 \\
I18306-0835 MM3 & $<$0.25 & --- &  --- & 0.32 (0.10)&& $<$0.31  & $<$0.36 \\
I18337-0743 MM1 & 1.06 (0.02)& 58.31 (0.04)& 3.51 (0.08)& 4.21 (0.09)&& $<$0.22  & $<$0.26 \\
I18337-0743 MM2 & 0.50 (0.03)& 58.13 (0.06)& 2.3 (0.1)& 1.22 (0.09)&& $<$0.24  & $<$0.28 \\
I18337-0743 MM3 & 0.25 (0.02)& 55.6 (0.2)& 3.7 (0.4)& 1.07 (0.09)&& $<$0.28  & $<$0.33 \\
\enddata 
\tablenotetext{a}{The quoted errors denote one standard deviation.}
\tablenotetext{b}{Velocity range for integration is $\pm5$ km s$^{-1}$ centered at the $LSR$ velocity of N$_2$H$^+$ $J$=1--0 or NH$_3$ (1, 1).}
\label{tab:t5}
\end{deluxetable} 

\clearpage

\begin{deluxetable}{lrrrr} 
\tablecolumns{5} 
\tablewidth{0pc}
\tabletypesize{\small}  
\tablecaption{Line parameters of the CH$_3$OH $J_K$=$7_0$--$6_0$ $A^+$ line.\tablenotemark{a}} 
\tablehead{ 
\colhead{Source}& \colhead{$T_{\rm pk}$} & \colhead{$V_{LSR}$}   & \colhead{$\Delta V$} & \colhead{$\int T_a^* dV$\tablenotemark{b}} \\
& [K] & [km s$^{-1}$] & [km s$^{-1}$] & [K km s$^{-1}$] \\
 }
\startdata 
G015.05+00.07 MM1 & $<$0.29  &  --- & --- & $<$0.6   \\
G015.31-00.16 MM2 & $<$0.30  & ---  & ---  & $<$0.6   \\
G015.31-00.16 MM3 & $<$0.40  & ---  & ---  & $<$0.8   \\
G019.27+0.07 MM1 & 0.57 (0.04)& 26.7 (0.2)& 5.2 (0.4)& 3.1 (0.2) \\
G019.27+0.07 MM2 & $<$0.28  &---   & ---  & $<$0.6   \\
G022.35+00.41 MM1 & 0.64 (0.04)& 52.6 (0.2)& 5.4 (0.4)& 3.6 (0.2) \\
G022.35+00.41 MM2 & $<$0.34  &  --- &---   & $<$0.7   \\
G022.35+00.41 MM3 & $<$0.28  & ---  & ---  & $<$0.6   \\
G023.60+00.00 MM1 & 0.67 (0.04)& 107.1 (0.2)& 4.9 (0.4)& 3.5 (0.2) \\
G023.60+00.00 MM2 & 0.34 (0.03)& 53.0 (0.3)& 6.7 (0.7)& 2.2 (0.2) \\
G023.60+00.00 MM3 & $<$0.25  & ---  & ---  & $<$0.5   \\
G023.60+00.00 MM4 & $<$0.25  & ---  &---   & $<$0.5   \\
G023.60+00.00 MM7 & $<$0.39  &  --- &  --- & $<$0.8   \\
G024.08+00.04 MM1 & $<$0.34  & ---  & ---  & $<$0.7   \\
G024.08+00.04 MM2(1) & $<$0.46  & ---  & ---  & $<$1.0   \\
G024.08+00.04 MM2(2) & $<$0.46  & ---  & ---  & $<$1.0   \\
G024.08+00.04 MM3 & $<$0.31  & ---  & ---  & $<$0.7   \\
G024.08+00.04 MM4 & $<$0.39  & ---  & ---  & $<$0.8   \\
G024.33+00.11 MM1 & 0.48 (0.04)& 114.2 (0.3)& 6.6 (0.6)& 3.0 (0.2) \\
G024.33+00.11 MM2 & $<$0.29  & ---  & ---  & $<$0.6   \\
G024.33+00.11 MM3 & $<$0.19  & ---  & ---  & $<$0.4   \\
G024.33+00.11 MM4 & $<$0.28  & ---  & ---  & $<$0.6   \\
G024.33+00.11 MM5 & $<$0.28  & ---  & ---  & $<$0.6   \\
G024.33+00.11 MM6 & $<$0.36  & ---  & ---  & $<$0.8   \\
G024.33+00.11 MM8 & $<$0.24  & ---  & ---  & $<$0.5   \\
G024.33+00.11 MM9 & $<$0.30  & ---  & ---  & $<$0.6   \\
G024.33+00.11 MM11 & $<$0.26  & ---  & ---  & $<$0.5   \\
G024.60+00.08 MM1 & 0.32 (0.03)& 52.1 (0.3)& 5.9 (0.7)& 1.9 (0.2) \\
G024.60+00.08 MM2 & $<$0.32  & ---  &---   & $<$0.7   \\
G025.04-00.20 MM1 & $<$0.28  & ---  &---   & $<$0.6   \\
G025.04-00.20 MM2 & $<$0.27  & ---  &---   & $<$0.6   \\
G025.04-00.20 MM4 & $<$0.33  & ---  &  --- & $<$0.7   \\
G034.43+00.24 MM1 & 1.79 (0.09)& 57.6 (0.1)& 4.8 (0.3)& 9.5 (0.1) \\
G034.43+00.24 MM2 & 1.26 (0.06)& 57.7 (0.1)& 5.7 (0.3)& 7.3 (0.2) \\
G034.43+00.24 MM3 & 0.73 (0.03)& 58.8 (0.2)& 8.6 (0.5)& 5.4 (0.1) \\
G034.43+00.24 MM4 & 0.57 (0.05)& 57.2 (0.2)& 4.1 (0.4)& 2.5 (0.2) \\
G034.43+00.24 MM5 & $<$0.30  & ---  &  --- & $<$0.6   \\
G034.43+00.24 MM6 & $<$0.31  & ---  & ---  & $<$0.7   \\
G034.43+00.24 MM8 & $<$0.28  & ---  & ---  & $<$0.6   \\
G034.43+00.24 MM9 & $<$0.32  & ---  & ---  & $<$0.7   \\
I18102-1800 MM1 & 0.67 (0.03)& 21.7 (0.2)& 6.2 (0.4)& 4.2 (0.1) \\
I18151-1208 MM1 & 0.30 (0.05)& 33.1 (0.3)& 3.6 (0.6)& 1.1 (0.2) \\
I18151-1208 MM2 & 0.51 (0.04)& 30.4 (0.3)& 7.1 (0.6)& 3.3 (0.2) \\
I18151-1208 MM3 & $<$0.27  & ---  & ---  & $<$0.6   \\
I18182-1433 MM1 & 0.43 (0.04)& 60.0 (0.2)& 4.4 (0.5)& 2.0 (0.2) \\
I18182-1433 MM2 & $<$0.39  & ---  & ---  & $<$0.8   \\
I18223-1243 MM1 & $<$0.37  & ---  & ---  & $<$0.8   \\
I18223-1243 MM2 & $<$0.42  & ---  & ---  & $<$0.9   \\
I18223-1243 MM3 & 0.27 (0.03)& 44.9 (0.3)& 5.4 (0.6)& 1.5 (0.1) \\
I18223-1243 MM4 & $<$0.29  &  &  & $<$0.6   \\
I18306-0835 MM1 & 0.50 (0.04)& 78.0 (0.2)& 3.6 (0.4)& 1.6 (0.1) \\
I18306-0835 MM2 & $<$0.45  & ---  & ---  & $<$0.9   \\
I18306-0835 MM3 & $<$0.43  & ---  & ---  & $<$0.9   \\
I18337-0743 MM1 & 0.34 (0.03)& 58.1 (0.3)& 5.7 (0.6)& 1.9 (0.1) \\
I18337-0743 MM2 & $<$0.25  & ---  & ---  & $<$0.5   \\
I18337-0743 MM3 & 0.23 (0.03)& 56.1 (0.3)& 5.0 (0.8)& 1.3 (0.1) \\
\enddata 
\tablenotetext{a}{The quoted errors denote one standard deviation.}
\tablenotetext{b}{Velocity range for integration is $\pm5$ km s$^{-1}$ centered at the $LSR$ velocity of N$_2$H$^+$ $J$=1--0 or NH$_3$ (1, 1).}
\label{tab:t6}
\end{deluxetable} 

\clearpage

\begin{deluxetable}{lrrrr} 
\tablecolumns{5} 
\tablewidth{0pc} 
\tabletypesize{\small} 
\tablecaption{Line parameters of the NH$_3$ ($J$, $K$) = (1, 1) line.\tablenotemark{a}} 
\tablehead{ 
\colhead{Source}& \colhead{$T_{\rm pk}$} & \colhead{$V_{LSR}$}   & \colhead{$\Delta V$} & \colhead{$\int T_a^* dV$\tablenotemark{b}} \\
& [K] & [km s$^{-1}$] & [km s$^{-1}$] & [K km s$^{-1}$] \\
 }
\startdata 
G015.05+00.07 MM1 & 0.41 (0.02) & 24.72 (0.04) & 2.3 (0.1) & 1.16  (0.04) \\
G015.31-00.16 MM2 & 0.45 (0.02) & 31.05 (0.03) & 1.34 (0.07) & 0.72  (0.04) \\
G015.31-00.16 MM3 & 0.38 (0.02) & 30.93 (0.03) & 1.27 (0.07) & 0.50  (0.04) \\
G019.27+0.07 MM1 & 0.93 (0.03) & 26.37 (0.03) & 2.57 (0.08) & 2.85  (0.04) \\
G019.27+0.07 MM2 & 0.86 (0.02) & 26.79 (0.04) & 2.71 (0.09) & 2.68  (0.04) \\
G022.35+00.41 MM1 & 0.36 (0.01) & 52.67 (0.05) & 3.0 (0.1) & 1.31  (0.03) \\
G022.35+00.41 MM2 & 0.29 (0.02) & 60.22 (0.04) & 1.5 (0.1) & 0.53  (0.04) \\
G022.35+00.41 MM3 & 0.78 (0.02) & 84.29 (0.03) & 2.45 (0.07) & 2.06  (0.04) \\
G023.60+00.00 MM1 & 0.60 (0.02) & 106.10 (0.05) & 4.1 (0.1) & 2.72  (0.03) \\
G023.60+00.00 MM2 & 1.02 (0.02) & 53.49 (0.03) & 2.47 (0.07) & 2.93  (0.04) \\
G023.60+00.00 MM3 & 0.76 (0.02) & 105.45 (0.05) & 3.6 (0.1) & 3.06  (0.04) \\
G023.60+00.00 MM4 & 0.78 (0.02) & 53.65 (0.03) & 2.2 (0.07) & 2.06  (0.04) \\
G023.60+00.00 MM7 & 1.04 (0.03) & 53.74 (0.03) & 2.22 (0.07) & 2.74  (0.04) \\
G024.08+00.04 MM1 & 0.55 (0.02) & 113.61 (0.04) & 2.9 (0.1) & 1.92  (0.04) \\
G024.08+00.04 MM2(1) & 0.31 (0.02) & 51.80 (0.06) & 1.6 (0.1) & 0.69  (0.04) \\
G024.08+00.04 MM2(2) & 0.39 (0.02) & 114.02 (0.05) & 2.4 (0.1) & 1.14  (0.04) \\
G024.08+00.04 MM3 & 0.52 (0.02) & 51.59 (0.03) & 1.89 (0.07) & 1.21  (0.04) \\
G024.08+00.04 MM4 & 0.55 (0.02) & 51.59 (0.03) & 1.67 (0.07) & 1.18  (0.04) \\
G024.33+00.11 MM1 & 1.25 (0.04) & 113.61 (0.05) & 3.4 (0.1) & 4.88  (0.06) \\
G024.33+00.11 MM2 & 0.71 (0.02) & 118.11 (0.05) & 3.5 (0.1) & 2.50  (0.05) \\
G024.33+00.11 MM3 p1& 0.58 (0.01) & 116.63 (0.04) & 2.16 (0.09) & 2.26  (0.05) \\
G024.33+00.11 MM3 p2 & 0.47 (0.01) & 119.09 (0.05) & 1.9 (0.1) &   \\
G024.33+00.11 MM4 & 0.55 (0.02) & 114.89 (0.07) & 4.6 (0.2) & 2.84  (0.05) \\
G024.33+00.11 MM5 & 0.97 (0.01) & 117.47 (0.03) & 3.06 (0.06) & 3.18  (0.05) \\
G024.33+00.11 MM6 & 1.02 (0.03) & 113.85 (0.05) & 3.2 (0.1) & 3.77  (0.06) \\
G024.33+00.11 MM8 & 0.29 (0.02) & 119.93 (0.09) & 2.8 (0.2) & 1.65  (0.03) \\
G024.33+00.11 MM9 p1& 0.38 (0.01) & 116.18 (0.04) & 1.98 (0.09) & 1.99  (0.03) \\
G024.33+00.11 MM9 p2  & 0.45 (0.01) & 119.28 (0.03) & 1.90 (0.08) &   \\
G024.33+00.11 MM11 & 0.59 (0.02) & 112.99 (0.04) & 2.4 (0.1) & 1.60  (0.06) \\
G024.60+00.08 MM1 & 0.93 (0.02) & 52.99 (0.03) & 2.37 (0.07) & 2.61  (0.04) \\
G024.60+00.08 MM2 & 0.34 (0.01) & 115.11 (0.04) & 2.3 (0.1) & 0.92  (0.04) \\
G025.04-00.20 MM1 & 1.18 (0.03) & 63.53 (0.02) & 2.08 (0.06) & 2.87  (0.04) \\
G025.04-00.20 MM2 & 0.86 (0.02) & 63.44 (0.03) & 2.56 (0.08) & 2.48  (0.04) \\
G025.04-00.20 MM4 & 1.00 (0.03) & 63.55 (0.03) & 2.03 (0.06) & 2.31  (0.04) \\
G034.43+00.24 MM1 & 1.77 (0.04) & 57.79 (0.04) & 3.39 (0.09) & 6.83  (0.03) \\
G034.43+00.24 MM2 & 1.58 (0.04) & 57.50 (0.04) & 3.8 (0.1) & 6.70  (0.03) \\
G034.43+00.24 MM3 & 1.16 (0.03) & 58.98 (0.03) & 2.56 (0.07) & 3.62  (0.04) \\
G034.43+00.24 MM4 & 1.43 (0.03) & 57.50 (0.04) & 3.28 (0.09) & 5.32  (0.04) \\
G034.43+00.24 MM5 & 0.73 (0.03) & 57.57 (0.04) & 2.3 (0.1) & 1.97  (0.06) \\
G034.43+00.24 MM6 & 0.85 (0.03) & 58.31 (0.04) & 2.59 (0.09) & 2.55  (0.06) \\
G034.43+00.24 MM8 & 1.29 (0.03) & 57.84 (0.04) & 3.14 (0.09) & 4.62  (0.06) \\
G034.43+00.24 MM9 & 1.10 (0.03) & 58.53 (0.04) & 2.62 (0.09) & 3.28  (0.06) \\
I18102-1800 MM1 & 1.63 (0.04) & 21.53 (0.04) & 2.88 (0.08) & 5.61  (0.04) \\
I18151-1208 MM1 & 0.58 (0.02) & 32.84 (0.03) & 2.37 (0.08) & 1.59  (0.04) \\
I18151-1208 MM2 & 0.45 (0.01) & 29.93 (0.04) & 2.9 (0.1) & 1.48  (0.04) \\
I18151-1208 MM3 & 0.23 (0.01) & 30.86 (0.06) & 2.1 (0.1) & 0.42  (0.04) \\
I18182-1433 MM1 & 0.75 (0.02) & 59.72 (0.04) & 3.18 (0.09) & 2.59  (0.04) \\
I18182-1433 MM2 & 0.53 (0.02) & 40.93 (0.04) & 2.08 (0.09) & 1.22  (0.05) \\
I18223-1243 MM1 & 0.93 (0.03) & 45.28 (0.03) & 1.87 (0.07) & 3.03  (0.03) \\
I18223-1243 MM2 & 0.82 (0.03) & 45.84 (0.05) & 2.9 (0.1) & 2.16  (0.06) \\
I18223-1243 MM3 & 0.92 (0.02) & 45.66 (0.04) & 2.95 (0.09) & 2.62  (0.06) \\
I18223-1243 MM4 & 0.62 (0.02) & 45.78 (0.03) & 1.73 (0.08) & 1.34  (0.06) \\
I18306-0835 MM1 & 0.91 (0.02) & 77.88 (0.03) & 2.57 (0.08) & 2.75  (0.04) \\
I18306-0835 MM2 & 0.29 (0.02) & 76.58 (0.07) & 2.3 (0.2) & 0.82  (0.06) \\
I18306-0835 MM3 & 0.37 (0.02) & 54.45 (0.06) & 2.1 (0.1) & 0.73  (0.06) \\
I18337-0743 MM1 & 1.38 (0.04) & 58.37 (0.05) & 3.7 (0.1) & 5.68  (0.04) \\
I18337-0743 MM2 & 0.32 (0.01) & 58.53 (0.07) & 3.2 (0.2) & 1.15  (0.04) \\
I18337-0743 MM3 & 0.45 (0.02) & 55.85 (0.05) & 3.2 (0.1) & 1.67  (0.05) \\
\enddata 
\tablenotetext{a}{The quoted errors denote one standard deviation.}
\tablenotetext{b}{Velocity range for integration is $\pm5$ km s$^{-1}$ centered at the $LSR$ velocity of N$_2$H$^+$ $J$=1--0 or NH$_3$ (1, 1).}
\label{tab:t7}
\end{deluxetable}

\clearpage

\begin{deluxetable}{lrrrr} 
\tablecolumns{5} 
\tablewidth{0pc} 
\tabletypesize{\small} 
\tablecaption{Line parameters of the NH$_3$ ($J$, $K$) = (2, 2) line.\tablenotemark{a}} 
\tablehead{ 
\colhead{Source}& \colhead{$T_{\rm pk}$} & \colhead{$V_{LSR}$}   & \colhead{$\Delta V$} & \colhead{$\int T_a^* dV$\tablenotemark{b}} \\
& [K] & [km s$^{-1}$] & [km s$^{-1}$] & [K km s$^{-1}$] \\
 }
\startdata 
G015.05+00.07 MM1 & 0.18 (0.01) & 24.60 (0.08) & 2.1 (0.2) & 0.48 (0.05) \\
G015.31-00.16 MM2 & $<$0.09 & --- & --- & $<$0.14 \\
G015.31-00.16 MM3 & $<$0.08 & --- & --- & $<$0.13 \\
G019.27+0.07 MM1 & 0.43 (0.01) & 26.37 (0.03) & 2.24 (0.07) & 1.08 (0.04) \\
G019.27+0.07 MM2 & 0.40 (0.01) & 26.92 (0.04) & 2.44 (0.09) & 1.19 (0.04) \\
G022.35+00.41 MM1 & 0.17 (0.01) & 52.57 (0.07) & 2.0 (0.2) & 0.43 (0.04) \\
G022.35+00.41 MM2 & $<$0.08 & --- & --- & 0.27 (0.05) \\
G022.35+00.41 MM3 & 0.30 (0.01) & 84.32 (0.03) & 2.1 (0.1) & 0.81 (0.05) \\
G023.60+00.00 MM1 & 0.31 (0.01) & 106.37 (0.04) & 3.3 (0.1) & 1.11 (0.03) \\
G023.60+00.00 MM2 & 0.46 (0.01) & 53.47 (0.03) & 2.62 (0.07) & 1.47 (0.04) \\
G023.60+00.00 MM3 & 0.32 (0.01) & 105.73 (0.06) & 3.8 (0.1) & 1.29 (0.04) \\
G023.60+00.00 MM4 & 0.30 (0.01) & 53.44 (0.04) & 2.2 (0.1) & 0.71 (0.04) \\
G023.60+00.00 MM7 & 0.33 (0.01) & 53.6 (0.04) & 2.0 (0.1) & 0.73 (0.05) \\
G024.08+00.04 MM1 & 0.30 (0.01) & 113.66 (0.05) & 2.5 (0.1) & 0.87 (0.04) \\
G024.08+00.04 MM2(1) & 0.12 (0.02) & 51.75 (0.08) & 1.1 (0.2) & 0.12 (0.04) \\
G024.08+00.04 MM2(2) & 0.17 (0.01) & 114.17 (0.07) & 2.1 (0.2) & 0.38 (0.04) \\
G024.08+00.04 MM3 & 0.12 (0.01) & 51.62 (0.09) & 1.7 (0.2) & 0.16 (0.04) \\
G024.08+00.04 MM4 & 0.15 (0.01) & 51.63 (0.07) & 1.5 (0.2) & 0.23 (0.04) \\
G024.33+00.11 MM1 & 0.84 (0.02) & 113.46 (0.03) & 3.04 (0.07) & 2.84 (0.06) \\
G024.33+00.11 MM2 & 0.34 (0.01) & 118.00 (0.06) & 3.1 (0.2) & 1.05 (0.06) \\
G024.33+00.11 MM3 p1 & 0.18 (0.01) & 116.9 (0.1) & 2.4 (0.3) & 0.62 (0.06) \\
G024.33+00.11 MM3 p2 & 0.17 (0.02) & 119.24 (0.08) & 1.2 (0.2) & \\
G024.33+00.11 MM4 & 0.16 (0.01) & 114.61 (0.2) & 4.8 (0.4) & 0.80 (0.05) \\
G024.33+00.11 MM5 & 0.47 (0.01) & 117.4 (0.05) & 2.9 (0.1) & 1.28 (0.05) \\
G024.33+00.11 MM6 & 0.62 (0.02) & 113.76 (0.04) & 3.02 (0.09) & 2.08 (0.06) \\
G024.33+00.11 MM8 & 0.10 (0.01) & 119.4 (0.1) & 2.0 (0.3) & 0.41 (0.03) \\
G024.33+00.11 MM9 p1 & 0.13 (0.01) & 116.26 (0.07) & 2.2 (0.2) & 0.62 (0.03) \\
G024.33+00.11 MM9 p2 & 0.19 (0.01) & 119.22 (0.04) & 1.4 (0.1) &  \\
G024.33+00.11 MM11 & 0.24 (0.02) & 112.91 (0.09) & 2.5 (0.2) & 0.67 (0.06) \\
G024.60+00.08 MM1 & 0.37 (0.01) & 52.98 (0.04) & 2.4 (0.1) & 1.10 (0.04) \\
G024.60+00.08 MM2 & 0.14 (0.01) & 114.7 (0.1) & 3.4 (0.3) & 0.55 (0.04) \\
G025.04-00.20 MM1 & 0.44 (0.01) & 63.6 (0.03) & 2.18 (0.08) & 1.12 (0.04) \\
G025.04-00.20 MM2 & 0.25 (0.01) & 63.52 (0.07) & 2.9 (0.2) & 0.88 (0.05) \\
G025.04-00.20 MM4 & 0.28 (0.01) & 63.61 (0.05) & 2.1 (0.1) & 0.64 (0.05) \\
G034.43+00.24 MM1 & 1.02 (0.01) & 57.71 (0.02) & 3.54 (0.04) & 4.06 (0.04) \\
G034.43+00.24 MM2 & 0.93 (0.01) & 57.53 (0.02) & 3.66 (0.05) & 3.73 (0.04) \\
G034.43+00.24 MM3 & 0.46 (0.01) & 58.97 (0.03) & 2.91 (0.08) & 1.60 (0.04) \\
G034.43+00.24 MM4 & 0.76 (0.01) & 57.49 (0.02) & 3.17 (0.05) & 2.64 (0.04) \\
G034.43+00.24 MM5 & 0.28 (0.02) & 57.59 (0.07) & 1.9 (0.2) & 0.57 (0.06) \\
G034.43+00.24 MM6 & 0.27 (0.02) & 58.62 (0.08) & 2.5 (0.2) & 0.85 (0.07) \\
G034.43+00.24 MM8 & 0.63 (0.02) & 57.94 (0.04) & 3.3 (0.1) & 2.37 (0.07) \\
G034.43+00.24 MM9 & 0.36 (0.02) & 58.99 (0.06) & 2.3 (0.1) & 0.82 (0.07) \\
I18102-1800 MM1 & 0.93 (0.01) & 21.57 (0.02) & 3.04 (0.05) & 3.28 (0.05) \\
I18151-1208 MM1 & 0.34 (0.01) & 32.92 (0.04) & 2.2 (0.1) & 0.90 (0.05) \\
I18151-1208 MM2 & 0.20 (0.01) & 30.0 (0.1) & 4.2 (0.2) & 0.97 (0.05) \\
I18151-1208 MM3 & 0.11 (0.01) & 31.2 (0.2) & 3.6 (0.4) & 0.45 (0.04) \\
I18182-1433 MM1 & 0.39 (0.01) & 59.69 (0.04) & 3.0 (0.1) & 1.15 (0.04) \\
I18182-1433 MM2 & 0.17 (0.01) & 40.9 (0.1) & 2.6 (0.2) & 0.42 (0.05) \\
I18223-1243 MM1 & 0.48 (0.02) & 45.20 (0.04) & 2.2 (0.1) & 1.21 (0.04) \\
I18223-1243 MM2 & 0.33 (0.02) & 45.72 (0.07) & 2.8 (0.2) & 1.28 (0.06) \\
I18223-1243 MM3 & 0.42 (0.01) & 45.69 (0.03) & 2.57 (0.07) & 0.93 (0.06) \\
I18223-1243 MM4 & 0.23 (0.02) & 45.75 (0.08) & 1.8 (0.2) & 0.56 (0.06) \\
I18306-0835 MM1 & 0.49 (0.01) & 77.87 (0.04) & 2.82 (0.08) & 1.56 (0.04) \\
I18306-0835 MM2 & 0.12 (0.02) & 76.7 (0.2) & 2.5 (0.4) & 0.23 (0.06) \\
I18306-0835 MM3 & 0.13 (0.02) & 55.2 (0.2) & 2.3 (0.4) & 0.30 (0.06) \\
I18337-0743 MM1 & 0.77 (0.01) & 58.37 (0.03) & 3.52 (0.06) & 2.92 (0.05) \\
I18337-0743 MM2 & 0.18 (0.01) & 59.99 (0.09) & 2.4 (0.2) & 0.37 (0.05) \\
I18337-0743 MM3 & 0.16 (0.01) & 56.3 (0.1) & 2.9 (0.3) & 0.59 (0.05) \\
\enddata 
\tablenotetext{a}{The quoted errors denote one standard deviation.}
\tablenotetext{b}{Velocity range for integration is $\pm5$ km s$^{-1}$ centered at the $LSR$ velocity of N$_2$H$^+$ $J$=1--0 or NH$_3$ (1, 1).}
\label{tab:t8}
\end{deluxetable}

\clearpage

\begin{deluxetable}{lrrrr} 
\tablecolumns{5} 
\tablewidth{0pc} 
\tabletypesize{\small} 
\tablecaption{Line parameters of the NH$_3$ ($J$, $K$) = (3, 3) line.\tablenotemark{a}} 
\tablehead{ 
\colhead{Source}& \colhead{$T_{\rm pk}$} & \colhead{$V_{LSR}$}   & \colhead{$\Delta V$} & \colhead{$\int T_a^* dV$\tablenotemark{b}} \\
& [K] & [km s$^{-1}$] & [km s$^{-1}$] & [K km s$^{-1}$] \\
 }
\startdata 
G015.05+00.07 MM1 & $<$0.08  & --- & --- & 0.34 (0.04) \\
G015.31-00.16 MM2 & $<$0.08  &---    &  --- & $<$0.13   \\
G015.31-00.16 MM3 & $<$0.08  & ---  &  ---  & $<$0.12   \\
G019.27+0.07 MM1 &  0.16 (0.01) & 26.6 (0.1) & 4.0 (0.3) & 0.69 (0.04) \\
G019.27+0.07 MM2 &  0.16 (0.01) & 27.0 (0.1) & 3.9 (0.3) & 0.77 (0.04) \\
G022.35+00.41 MM1 &  0.15 (0.01) & 52.97 (0.07) & 4.5 (0.2) & 0.86 (0.04) \\
G022.35+00.41 MM2 & $<$0.08  &  --- &  --- & $<$0.12   \\
G022.35+00.41 MM3 & $<$0.08  &  ---& --- & 0.24 (0.04) \\
G023.60+00.00 MM1 &  0.19 (0.01) & 106.56 (0.09) & 5.0 (0.2) & 0.99 (0.03) \\
G023.60+00.00 MM2 &  0.23 (0.01) & 53.51 (0.08) & 5.2 (0.2) & 1.29 (0.03) \\
G023.60+00.00 MM3 &  0.17 (0.01) & 106.3 (0.1) & 4.3 (0.3) & 0.82 (0.04) \\
G023.60+00.00 MM4 &  0.19 (0.01) & 53.46 (0.09) & 2.9 (0.2) & 0.72 (0.05) \\
G023.60+00.00 MM7 &  0.14 (0.01) & 53.3 (0.1) & 4.2 (0.3) & 0.72 (0.04) \\
G024.08+00.04 MM1 &  0.11 (0.01) & 113.9 (0.1) & 2.7 (0.3) & 0.28 (0.03) \\
G024.08+00.04 MM2(1) & $<$0.07 &  ---&  --- & $<$0.11   \\
G024.08+00.04 MM2(2) & $<$0.07  &  --- &  --- & $<$0.11   \\
G024.08+00.04 MM3 & $<$0.07  &  --- & ---& 0.13 (0.04) \\
G024.08+00.04 MM4 & $<$0.07  &--- & ---& $<$0.11   \\
G024.33+00.11 MM1 &  0.70 (0.01) & 113.81 (0.04) & 4.2 (0.1) & 3.30 (0.06) \\
G024.33+00.11 MM2 &  0.18 (0.01) & 117.6 (0.2) & 5.6 (0.4) & 1.01 (0.05) \\
G024.33+00.11 MM3 &  0.12 (0.01) & 118.2 (0.2) & 5.7 (0.5) & 0.76 (0.05) \\
G024.33+00.11 MM4 &  0.14 (0.02) & 114.9 (0.1) & 2.1 (0.3) & 0.31 (0.05) \\
G024.33+00.11 MM5 &  0.19 (0.01) & 119.1 (0.1) & 5.0 (0.3) & 1.02 (0.05) \\
G024.33+00.11 MM6 &  0.48 (0.01) & 113.98 (0.05) & 3.7 (0.1) & 2.07 (0.06) \\
G024.33+00.11 MM8 & $<$0.07 & ---  &  --- & 0.19 (0.03) \\
G024.33+00.11 MM9 & $<$0.07 &  --- &  --- & 0.26 (0.03) \\
G024.33+00.11 MM11 & $<$0.12 & --- & --- & 0.36 (0.06) \\
G024.60+00.08 MM1 &  0.11 (0.01) & 52.2 (0.2) & 7.6 (0.5) & 0.78 (0.04) \\
G024.60+00.08 MM2 &  $<$0.08 & --- & --- & 0.41 (0.04) \\
G025.04-00.20 MM1 &  0.13 (0.01) & 63.2 (0.1) & 3.0 (0.3) & 0.44 (0.04) \\
G025.04-00.20 MM2 &  0.10 (0.01) & 63.7 (0.2) & 3.4 (0.4) & 0.41 (0.04) \\
G025.04-00.20 MM4 & $<$0.08 &  --- &  --- & 0.18 (0.04) \\
G034.43+00.24 MM1 &  0.66 (0.01) & 57.9 (0.03) & 5.17 (0.08) & 3.58 (0.04) \\
G034.43+00.24 MM2 &  0.59 (0.01) & 57.65 (0.04) & 5.76 (0.09) & 3.45 (0.04) \\
G034.43+00.24 MM3 &  0.215 (0.006) & 59.4 (0.1) & 7.2 (0.2) & 1.42 (0.04) \\
G034.43+00.24 MM4 &  0.399 (0.007) & 57.55 (0.05) & 5.2 (0.1) & 2.15 (0.04) \\
G034.43+00.24 MM5 & $<$0.12 & ---& --- & 0.39 (0.06) \\
G034.43+00.24 MM6 & $<$0.12 &  --- &  --- & 0.33 (0.06) \\
G034.43+00.24 MM8 &  0.33 (0.02) & 57.8 (0.1) & 4.3 (0.2) & 1.53 (0.07) \\
G034.43+00.24 MM9 &  0.19 (0.02) & 59.2 (0.1) & 2.9 (0.3) & 0.64 (0.07) \\
I18102-1800 MM1 &  0.61 (0.01) & 21.47 (0.03) & 4.71 (0.08) & 3.01 (0.04) \\
I18151-1208 MM1 &  0.14 (0.01) & 32.8 (0.1) & 3.4 (0.3) & 0.58 (0.04) \\
I18151-1208 MM2 &  0.11 (0.01) & 29.4 (0.2) & 6.8 (0.5) & 0.72 (0.05) \\
I18151-1208 MM3 &  $<$0.08 & --- & --- & 0.29 (0.04) \\
I18182-1433 MM1 &  0.17 (0.01) & 59.4 (0.1) & 6 (0.3) & 1.02 (0.04) \\
I18182-1433 MM2 & $<$0.09 & ---& ---& 0.19 (0.05) \\
I18223-1243 MM1 &  0.16 (0.02) & 45.2 (0.1) & 1.9 (0.3) & 0.36 (0.06) \\
I18223-1243 MM2 & $<$0.12 & --- & --- & 0.32 (0.06) \\
I18223-1243 MM3 &  0.15 (0.01) & 45.1 (0.1) & 5.3 (0.3) & 0.8 (0.03) \\
I18223-1243 MM4 & $<$0.12 & ---& --- & 0.24 (0.07) \\
I18306-0835 MM1 &  0.31 (0.01) & 78.04 (0.06) & 3.5 (0.1) & 1.22 (0.04) \\
I18306-0835 MM2 &  0.09 (0.02) & 77.6 (0.2) & 2.8 (0.5) & 0.3 (0.06) \\
I18306-0835 MM3 & $<$0.11 & --- & --- & $<$0.18   \\
I18337-0743 MM1 &  0.39 (0.01) & 58.29 (0.06) & 4.8 (0.1) & 1.99 (0.05) \\
I18337-0743 MM2 &  0.13 (0.01) & 58.3 (0.1) & 2.5 (0.3) & 0.32 (0.04) \\
I18337-0743 MM3 &  $<$0.08 & --- & --- & 0.67 (0.04) \\
\enddata 
\tablenotetext{a}{The quoted errors denote one standard deviation.}
\tablenotetext{b}{Velocity range for integration is $\pm5$ km s$^{-1}$ centered at the $LSR$ velocity of N$_2$H$^+$ $J$=1--0 or NH$_3$ (1, 1).}
\label{tab:t9}
\end{deluxetable} 

\clearpage

\begin{deluxetable}{lrrrrrrrr} 
\rotate
\tablecolumns{9} 
\tablewidth{0pc} 
\tabletypesize{\scriptsize} 
\tablecaption{The NH$_3$ rotation temperatures and column densities.} 
\tablehead{ 
\colhead{Source}& \colhead{$T_{\rm rot}$(NH$_3$)} & \colhead{$N$(N$_2$H$^+$)\tablenotemark{a}}   & \colhead{$N$(CCS)\tablenotemark{a}} & \colhead{$N$(HC$_3$N)\tablenotemark{a}} &\colhead{$N$(NH$_3$)} & \colhead{$N$(CH$_3$OH)\tablenotemark{b}} & \colhead{$N$(CCS)/$N$(N$_2$H$^+$)}&\colhead{$N$(CCS)/$N$(NH$_3$)}  \\
& [K] & [10$^{12}$ cm$^{-2}$] & [10$^{12}$ cm$^{-2}$]& [10$^{12}$ cm$^{-2}$] & [10$^{15}$ cm$^{-2}$] & [10$^{14}$ cm$^{-5}$] &  & [10$^{-2}$] \\
 }
\startdata 
G015.05+00.07 MM1 & 15.2$_{-1.5}^{+1.7}$& 17.4$_{-1.0}^{+1.2}$&$<$3.8 & 9.7$_{-0.4}^{+0.5}$& 2.7$_{-0.8}^{+0.6}$& $<$2.2 &$<$0.24 &$<$0.20 \\
G015.31-00.16 MM2 & 10.3$_{-0.6}^{+0.7}$& 5.4$_{-0.1}^{+0.1}$&$<$1.6 & 5.0$_{-0.1}^{+0.1}$& 1.1$_{-0.6}^{+0.5}$& $<$2.3 &$<$0.30 &$<$0.32 \\
G015.31-00.16 MM3 & 12.6$_{-2.0}^{+2.4}$& 6.9$_{-0.5}^{+0.7}$&$<$1.7 & 4.4$_{-0.2}^{+0.3}$& 0.6$_{-0.5}^{+0.4}$& $<$3.1 &$<$0.27 &$<$1.74 \\
G019.27+0.07 MM1 & 15.4$_{-1.0}^{+1.1}$& 52.0$_{-2.1}^{+2.4}$&$<$3.7 & 24.2$_{-0.7}^{+0.8}$& 3.1$_{-0.6}^{+0.6}$& 7.9---16 &$<$0.07 &$<$0.15 \\
G019.27+0.07 MM2 & 15.8$_{-0.9}^{+1.0}$& 29.2$_{-1.1}^{+1.2}$&$<$3.5 & 22.1$_{-0.6}^{+0.7}$& 3.0$_{-0.6}^{+0.5}$& $<$2.2 &$<$0.12 &$<$0.15 \\
G022.35+00.41 MM1 & 16.0$_{-1.4}^{+1.6}$& 26.7$_{-1.4}^{+1.7}$&$<$3.6 & 18.3$_{-0.7}^{+0.8}$& 3.3$_{-0.8}^{+0.7}$& 9.2---18.7 &$<$0.14 &$<$0.15 \\
G022.35+00.41 MM2 & 12.8$_{-1.7}^{+2.2}$& 4.9$_{-0.3}^{+0.4}$&$<$1.7 & $<$2.2 & 1.3$_{-0.9}^{+0.6}$& $<$2.6 &$<$0.37 &$<$0.42 \\
G022.35+00.41 MM3 & 15.2$_{-1.0}^{+1.1}$& 15.8$_{-0.6}^{+0.7}$&$<$3.0 & 5.5$_{-0.1}^{+0.2}$& 2.2$_{-0.5}^{+0.5}$& $<$2.1 &$<$0.20 &$<$0.18 \\
G023.60+00.00 MM1 & 17.0$_{-1.2}^{+1.3}$& 36.7$_{-1.7}^{+1.9}$&$<$5.2 & 22.2$_{-0.7}^{+0.8}$& 4.3$_{-0.7}^{+0.7}$& 8.9---18 &$<$0.15 &$<$0.14 \\
G023.60+00.00 MM2 & 16.0$_{-0.9}^{+1.0}$& 22.8$_{-0.8}^{+0.9}$&$<$3.8 & 31.6$_{-0.8}^{+0.9}$& 2.5$_{-0.5}^{+0.5}$& 5.5---11.2 &$<$0.17 &$<$0.19 \\
G023.60+00.00 MM3 & 16.4$_{-1.1}^{+1.2}$& 17.5$_{-0.7}^{+0.8}$&$<$4.8 & 3.9$_{-0.1}^{+0.1}$& 2.6$_{-0.7}^{+0.6}$& $<$1.9 &$<$0.28 &$<$0.25 \\
G023.60+00.00 MM4 & 14.9$_{-0.9}^{+1.1}$& 8.6$_{-0.3}^{+0.4}$&$<$3.5 & 7.0$_{-0.2}^{+0.2}$& 2.1$_{-0.5}^{+0.5}$& $<$2.0 &$<$0.42 &$<$0.22 \\
G023.60+00.00 MM7 & 13.1$_{-0.7}^{+0.8}$& 16.0$_{-0.5}^{+0.5}$&$<$2.6 & 13.6$_{-0.2}^{+0.3}$& 2.7$_{-0.6}^{+0.5}$& $<$3.0 &$<$0.17 &$<$0.13 \\
G024.08+00.04 MM1 & 16.4$_{-1.4}^{+1.6}$& 38.9$_{-2.1}^{+2.4}$&$<$3.8 & 8.5$_{-0.3}^{+0.4}$& 4.0$_{-0.9}^{+0.7}$& $<$3.6 &$<$0.10 &$<$0.12 \\
G024.08+00.04 MM2(1) & 14.9$_{-2.6}^{+3.4}$& 8.1$_{-0.8}^{+1.1}$&$<$2.6 & 4.0$_{-0.3}^{+0.4}$& 1.3$_{-0.9}^{+0.7}$& $<$3.6 &$<$0.36 &$<$0.65 \\
G024.08+00.04 MM2(2) & 15.1$_{-1.6}^{+1.9}$& 10.5$_{-0.7}^{+0.8}$&$<$3.1 & 3.5$_{-0.1}^{+0.2}$& 2.8$_{-0.9}^{+0.7}$& $<$3.6 &$<$0.32 &$<$0.16 \\
G024.08+00.04 MM3 & 12.1$_{-0.9}^{+1.0}$& 5.1$_{-0.2}^{+0.2}$&$<$3.1 & 6.7$_{-0.1}^{+0.2}$& 2.0$_{-0.6}^{+0.5}$& $<$2.4 &$<$0.63 &$<$0.22 \\
G024.08+00.04 MM4 & 12.9$_{-0.9}^{+1.1}$& 4.6$_{-0.2}^{+0.2}$&$<$1.7 & 7.2$_{-0.1}^{+0.2}$& 1.7$_{-0.5}^{+0.4}$& $<$3.0 &$<$0.39 &$<$0.14 \\
G024.33+00.11 MM1 & 18.8$_{-1.6}^{+1.8}$& 63.3$_{-3.6}^{+4.1}$&$<$6.6 & 33.0$_{-1.4}^{+1.7}$& 5.0$_{-0.8}^{+0.7}$& 7.7---15.6 &$<$0.11 &$<$0.16 \\
G024.33+00.11 MM2 & 16.6$_{-1.2}^{+1.3}$& 20.8$_{-0.9}^{+1.1}$&$<$5.1 & 12.2$_{-0.4}^{+0.5}$& 3.4$_{-0.8}^{+0.7}$& $<$2.6 &$<$0.26 &$<$0.20 \\
G024.33+00.11 MM3 p1& 13.9$_{-0.7}^{+0.7}$& 21.9$_{-0.6}^{+0.6}$&$<$5.5 & 11.7$_{-0.2}^{+0.2}$& 2.1$_{-0.4}^{+0.3}$& $<$1.5 &$<$0.26 &$<$0.32 \\
G024.33+00.11 MM3 p2& 14.5$_{-1.3}^{+1.4}$&&&& 1.8$_{-0.4}^{+0.3}$& &&\\
G024.33+00.11 MM4 & 13.4$_{-0.9}^{+1.0}$& 11.5$_{-0.4}^{+0.5}$&$<$4.6 & 9.8$_{-0.2}^{+0.3}$& 4.2$_{-1.1}^{+0.9}$& $<$2.1 &$<$0.42 &$<$0.15 \\
G024.33+00.11 MM5 & 17.1$_{-0.5}^{+0.5}$& 23.7$_{-0.5}^{+0.5}$&$<$4.7 & 14.4$_{-0.2}^{+0.2}$& 2.7$_{-0.2}^{+0.2}$& $<$2.2 &$<$0.20 &$<$0.19 \\
G024.33+00.11 MM6 & 17.3$_{-1.4}^{+1.6}$& 14.3$_{-0.8}^{+0.9}$&$<$5.7 & 7.2$_{-0.3}^{+0.3}$& 4.7$_{-0.8}^{+0.7}$& $<$2.8 &$<$0.42 &$<$0.15 \\
G024.33+00.11 MM8 & ---&---&---&---& ---& $<$1.9 &---&---\\
G024.33+00.11 MM9 p1& 15.1$_{-1.1}^{+1.2}$&&& & &&\\
G024.33+00.11 MM9 p2 & 15.2$_{-0.9}^{+1.0}$& 7.7$_{-0.3}^{+0.3}$&$<$3.9 & 4.6$_{-0.1}^{+0.1}$& 2.2$_{-0.4}^{+0.3}$& $<$2.3 &$<$0.52 &$<$0.21 \\
G024.33+00.11 MM11 & 14.8$_{-1.4}^{+1.7}$& 8.6$_{-0.5}^{+0.6}$&$<$4.2 & $<$2.5 & 2.7$_{-0.8}^{+0.6}$& $<$2.0 &$<$0.51 &$<$0.22 \\
G024.60+00.08 MM1 & 15.3$_{-0.8}^{+0.9}$& 28.5$_{-0.9}^{+1.0}$&$<$3.2 & 24.0$_{-0.5}^{+0.6}$& 2.1$_{-0.5}^{+0.4}$& 4.9---9.9 &$<$0.12 &$<$0.20 \\
G024.60+00.08 MM2 & 14.2$_{-1.1}^{+1.2}$& 25.9$_{-1.1}^{+1.3}$&$<$4.2 & 11.1$_{-0.3}^{+0.4}$& 3.1$_{-0.6}^{+0.5}$& $<$3.0 &$<$0.17 &$<$0.17 \\
G025.04-00.20 MM1 & 14.7$_{-0.8}^{+0.9}$& 27.9$_{-0.9}^{+1.0}$&$<$3.6 & 15.2$_{-0.3}^{+0.4}$& 2.0$_{-0.4}^{+0.4}$& $<$2.5 &$<$0.13 &$<$0.22 \\
G025.04-00.20 MM2 & 13.1$_{-0.7}^{+0.8}$& 21.5$_{-0.6}^{+0.7}$&$<$4.5 & 10.5$_{-0.2}^{+0.2}$& 2.7$_{-0.6}^{+0.5}$& $<$2.1 &$<$0.21 &$<$0.21 \\
G025.04-00.20 MM4 & 13.5$_{-0.8}^{+0.9}$& 16.2$_{-0.5}^{+0.6}$&$<$3.3 & 5.9$_{-0.1}^{+0.1}$& 1.7$_{-0.5}^{+0.4}$& $<$2.5 &$<$0.21 &$<$0.27 \\
G034.43+00.24 MM1 & 18.5$_{-1.0}^{+1.1}$& 133.6$_{-5.0}^{+5.4}$&$<$3.9 & 54.3$_{-1.6}^{+1.7}$& 3.4$_{-0.6}^{+0.6}$& 24.2---49.2 &$<$0.03 &$<$0.14 \\
G034.43+00.24 MM2 & 18.8$_{-1.0}^{+1.1}$& 113.9$_{-4.2}^{+4.5}$&$<$4.5 & 52.0$_{-1.5}^{+1.6}$& 3.8$_{-0.6}^{+0.6}$& 18.7---37.9 &$<$0.04 &$<$0.14 \\
G034.43+00.24 MM3 & 15.5$_{-0.9}^{+0.9}$& 59.6$_{-2.0}^{+2.2}$&$<$3.2 & 30.6$_{-0.7}^{+0.8}$& 2.2$_{-0.5}^{+0.4}$& 13.9---28.1 &$<$0.06 &$<$0.19 \\
G034.43+00.24 MM4 & 17.6$_{-1.0}^{+1.1}$& 66.0$_{-2.4}^{+2.7}$&$<$3.4 & 29.9$_{-0.8}^{+0.9}$& 3.3$_{-0.6}^{+0.6}$& 6.4---13 &$<$0.05 &$<$0.13 \\
G034.43+00.24 MM5 & 14.3$_{-1.3}^{+1.5}$& 16.8$_{-0.9}^{+1.0}$&$<$3.5 & 6.1$_{-0.2}^{+0.3}$& 2.7$_{-0.8}^{+0.6}$& $<$2.3 &$<$0.22 &$<$0.18 \\
G034.43+00.24 MM6 & 14.0$_{-1.1}^{+1.3}$& 15.4$_{-0.7}^{+0.8}$&$<$3.2 & 11.2$_{-0.3}^{+0.4}$& 2.3$_{-0.7}^{+0.5}$& $<$2.4 &$<$0.22 &$<$0.20 \\
G034.43+00.24 MM8 & 17.2$_{-1.1}^{+1.2}$& 14.0$_{-0.6}^{+0.6}$&$<$4.4 & 10.7$_{-0.3}^{+0.4}$& 2.8$_{-0.6}^{+0.5}$& $<$2.1 &$<$0.33 &$<$0.20 \\
G034.43+00.24 MM9 & 13.9$_{-0.9}^{+1.1}$& 20.1$_{-0.8}^{+0.9}$&$<$3.2 & 11.7$_{-0.3}^{+0.3}$& 2.5$_{-0.6}^{+0.5}$& $<$2.4 &$<$0.16 &$<$0.17 \\
I18102-1800 MM1 & 17.9$_{-1.1}^{+1.2}$& 51.2$_{-2.1}^{+2.3}$&$<$6.6 & 41.2$_{-1.2}^{+1.4}$& 3.2$_{-0.6}^{+0.5}$& 10.7---21.7 &$<$0.13 &$<$0.25 \\
I18151-1208 MM1 & 20.8$_{-1.7}^{+1.8}$& 34.9$_{-2.1}^{+2.2}$&$<$5.3 & 15.0$_{-0.7}^{+0.8}$& 1.3$_{-0.5}^{+0.5}$& 2.9---5.9 &$<$0.16 &$<$0.66 \\
I18151-1208 MM2 & 17.8$_{-1.4}^{+1.5}$& 45.7$_{-2.4}^{+2.6}$&$<$5.8 & 9.3$_{-0.4}^{+0.4}$& 1.5$_{-0.6}^{+0.6}$& 8.4---17 &$<$0.13 &$<$0.65 \\
I18151-1208 MM3 & 16.0$_{-2.1}^{+2.7}$& 13.2$_{-1.1}^{+1.4}$&$<$2.8 & 3.6$_{-0.2}^{+0.3}$& 2.4$_{-1.0}^{+0.8}$& $<$2.1 &$<$0.23 &$<$0.20 \\
I18182-1433 MM1 & 19.0$_{-1.4}^{+1.5}$& 55.8$_{-2.7}^{+3.0}$&$<$5.8 & 31.6$_{-1.2}^{+1.3}$& 1.9$_{-0.6}^{+0.6}$& 5.1---10.3 &$<$0.11 &$<$0.45 \\
I18182-1433 MM2 & 13.8$_{-1.1}^{+1.2}$& 17.0$_{-0.7}^{+0.8}$&$<$3.8 & 3.4$_{-0.1}^{+0.1}$& 2.0$_{-0.6}^{+0.5}$& $<$3.0 &$<$0.23 &$<$0.27 \\
I18223-1243 MM1 & 17.5$_{-1.5}^{+1.6}$& 42.4$_{-2.3}^{+2.6}$&$<$3.8 & 10.7$_{-0.4}^{+0.5}$& 1.8$_{-0.4}^{+0.4}$& $<$3.3 &$<$0.09 &$<$0.27 \\
I18223-1243 MM2 & 15.1$_{-1.3}^{+1.4}$& 25.2$_{-1.2}^{+1.5}$&$<$3.8 & 6.3$_{-0.2}^{+0.3}$& 2.9$_{-0.7}^{+0.6}$& $<$3.3 &$<$0.16 &$<$0.17 \\
I18223-1243 MM3 & 16.2$_{-0.9}^{+1.0}$& 43.9$_{-1.6}^{+1.7}$&$<$4.3 & 15.1$_{-0.4}^{+0.4}$& 2.9$_{-0.6}^{+0.5}$& 3.8---7.7 &$<$0.10 &$<$0.19 \\
I18223-1243 MM4 & 15.5$_{-1.5}^{+1.7}$& 14.6$_{-0.9}^{+1.0}$&$<$3.3 & 6.6$_{-0.3}^{+0.3}$& 1.2$_{-0.5}^{+0.4}$& $<$2.2 &$<$0.24 &$<$0.47 \\
I18306-0835 MM1 & 17.3$_{-1.1}^{+1.2}$& 22.6$_{-0.9}^{+1.0}$&$<$5.9 & 16.6$_{-0.5}^{+0.6}$& 2.8$_{-0.5}^{+0.5}$& 4.1---8.4 &$<$0.27 &$<$0.26 \\
I18306-0835 MM2 & 16.4$_{-2.9}^{+3.8}$& 10.4$_{-1.2}^{+1.6}$&$<$5.6 & 4.0$_{-0.3}^{+0.4}$& 1.6$_{-1.2}^{+0.9}$& $<$3.5 &$<$0.60 &$<$1.39 \\
I18306-0835 MM3 & 13.4$_{-1.7}^{+2.2}$& 4.3$_{-0.3}^{+0.4}$&$<$4.2 & 2.8$_{-0.1}^{+0.2}$& 2.7$_{-1.1}^{+0.7}$& $<$3.3 &$<$1.05 &$<$0.26 \\
I18337-0743 MM1 & 17.1$_{-0.9}^{+1.0}$& 67.2$_{-2.4}^{+2.6}$&$<$4.7 & 41.1$_{-1.1}^{+1.2}$& 4.6$_{-0.7}^{+0.6}$& 4.8---9.8 &$<$0.07 &$<$0.12 \\
I18337-0743 MM2 & 17.1$_{-1.8}^{+2.2}$& 21.5$_{-1.5}^{+1.8}$&$<$4.6 & 11.9$_{-0.6}^{+0.8}$& 4.1$_{-1.2}^{+1.0}$& $<$1.9 &$<$0.23 &$<$0.16 \\
I18337-0743 MM3 & 15.0$_{-1.4}^{+1.6}$& 19.1$_{-1.1}^{+1.3}$&$<$4.6 & 9.8$_{-0.4}^{+0.5}$& 2.5$_{-0.9}^{+0.8}$& 3.2---6.5 &$<$0.25 &$<$0.29 \\\enddata 
\tablenotetext{a}{T$_{\rm ex}$(N$_2$H$^+$) is assumed to be equal to $T_{\rm ex}$(NH$_3$). The quoted error is dominated by the error in $T_{\rm ex}$(NH$_3$).}
\tablenotetext{b}{T$_{\rm ex}$(CH$_3$OH) is assumed to be from 20 K to 50 K.}
\label{tab:t10}
\end{deluxetable} 

\clearpage

\begin{deluxetable}{lrrrrr} 
\tablecolumns{6} 
\tablewidth{0pc} 
\tablecaption{Integrated Intensities and rotation temperature of the CH$_3$OH $J$=7--6 line.\tablenotemark{a}} 
\tablehead{ 
\colhead{Source}& \colhead{$7_{-1}$--$6_{-1}$ $E$} & \colhead{$7_{1}$--$6_{1}$ $E$}   & \colhead{$7_{2}$--$6_{2}$ $E$} & \colhead{$7_{0}$--$6_{0}$ $E$} &\colhead{$T_{\rm rot}$}\\
& [K km s$^{-1}$] & [K km s$^{-1}$] &  [K km s$^{-1}$] & [K km s$^{-1}$] & [K] \\
 }
\startdata 
G019.27+0.07 MM1 & 3.2 (0.2)& $<$0.4 & 0.7 (0.1)& 1.7 (0.2)& 7.3 (0.8)\\
G022.35+00.41 MM1 & 2.9 (0.2)& 0.5 (0.1)& 0.7 (0.1)& 1.1 (0.2)& 8.4 (0.3)\\
G023.60+00.00 MM1 & 3.6 (0.2)& $<$0.4 & 0.5 (0.1)& 0.9 (0.2)& 6.9 (0.6)\\
G023.60+00.01 MM2 & 2.4 (0.2)& $<$0.3& $<$0.3& $<$0.5 & $<$7.4 \\
G024.33+00.11 MM1 & 2.9 (0.2)& $<$0.4 & 1.0 (0.1)& 2.3 (0.2)& 8.3 (2.4)\\
G024.60+00.08 MM1 & 1.9 (0.2)& $<$0.4 & 0.5 (0.1)& 0.6 (0.2)& 9.2 (1.1)\\
G034.43+00.24 MM1 & 8.7 (0.1)& 2.5 (0.1)& 3.4 (0.1)& 5.4 (0.1)& 10.9 (1.2)\\
G034.43+00.25 MM2 & 6.8 (0.2)& 1.2 (0.1)& 1.6 (0.1)& 3.1 (0.2)& 8.3 (0.5)\\
G034.43+00.26 MM3 & 6.3 (0.1)& 0.7 (0.1)& 1.2 (0.1)& 2.7 (0.1)& 7.0 (0.5)\\
G034.43+00.27 MM4 & 2.8 (0.2)& $<$0.4 & 0.7 (0.1)& $<$0.5 & 9.3 (3.1)\\
I18102-1800 MM1 & 4.5 (0.1)& 0.4 (0.1)& 1.0 (0.1)& 1.5 (0.1)& 7.3 (1.0)\\
I18151-1208 MM1 & 1.6 (0.2)& $<$0.4 & 0.8 (0.1)& 0.9 (0.2)& 12.1 (0.8)\\
I18151-1209 MM2 & 4.3 (0.2)& $<$0.4& 1.8 (0.1)& 2.7 (0.2)& 8.1 (2.8)\\
I18182-1433 MM1 & 2.0 (0.2)& 0.8 (0.1)& 0.8 (0.1)& 1.5 (0.2)& 12.4 (3.5)\\
I18223-1245 MM3 & 1.3 (0.1)& 0.3 (0.1)& $<$0.3 & 0.4 (0.1)& 9.5 (2.5)\\
I18306-0835 MM1 & 1.8 (0.2)& 0.5 (0.1)& 0.6 (0.1)& 0.9 (0.2)& 10.2 (1.1)\\
I18337-0743 MM1 & 1.8 (0.1)& $<$0.4 & $<$0.4 & 0.5 (0.1)& 8.6 (1.9)\\
I18337-0745 MM3 & 1.7 (0.1)& 0.3 (0.1)& $<$0.3 & 0.6 (0.1)& 8.2 (1.2)\\
\enddata 
\tablenotetext{a}{The quoted errors denote one standard deviation. Velocity range for integration is $\pm5$ km s$^{-1}$ centered at the $LSR$ velocity of N$_2$H$^+$ $J$=1--0 or NH$_3$ (1, 1).}
\label{tab:t11}
\end{deluxetable} 

\clearpage

\end{document}